\documentclass[useAMS,usenatbib,onecolumn]{mnras}
\usepackage{amsmath,bm,amssymb}
\usepackage{dcolumn}
\usepackage[utf8]{inputenc}
\usepackage{graphics,graphicx}
\usepackage{float}
\usepackage{threeparttable}
\usepackage{url}
\usepackage{epstopdf}
\usepackage{soul}

\usepackage{booktabs}
\usepackage{multirow}

\usepackage{color}

\newcommand{\X}{$X\,{}^{2}\Sigma^{+}$}
\newcommand{\B}{$B\,{}^{2}\Sigma^{+}$}
\newcommand{\Bp}{$B^{\prime}\,{}^{2}\Sigma^{+}$}
\newcommand{\BBp}{$B/B^{\prime}\,{}^{2}\Sigma^{+}$}
\newcommand{\A}{$A\,{}^{2}\Pi$}

\newcommand{\C}{$C\,{}^{2}\Sigma^{+}$}
\newcommand{\D}{$D\,{}^{2}\Sigma^{+}$}
\newcommand{\E}{$E\,{}^{2}\Pi$}

\newcommand{\Duo}{{\sc Duo}}

\title[ExoMol line lists -- XLV. CaH \& MgH]{ExoMol line lists -- XLV. Rovibronic molecular line lists of calcium monohydride (CaH) and magnesium monohydride (MgH)}

\date{\today}

\author[A. Owens et al.]
{Alec Owens,\thanks{The corresponding author: alec.owens.13@ucl.ac.uk} Sophie Dooley, Luke McLaughlin, Brandon Tan, Guanming Zhang,
\newauthor{Sergei N. Yurchenko\thanks{The corresponding author: s.yurchenko@ucl.ac.uk} and Jonathan Tennyson\thanks{The corresponding author: j.tennyson@ucl.ac.uk}}\vspace*{4mm}\\
Department of Physics and Astronomy, University College London, Gower Street, WC1E 6BT London, UK}
\date{Accepted XXXX. Received XXXX; in original form XXXX}

\pagerange{\pageref{firstpage}--\pageref{lastpage}} \pubyear{2021}

\begin{document}

\label{firstpage}

\maketitle

\begin{abstract}
New molecular line lists for calcium monohydride ($^{40}$Ca$^{1}$H) and magnesium monohydride ($^{24}$Mg$^{1}$H) and its minor isotopologues ($^{25}$Mg$^{1}$H and $^{26}$Mg$^{1}$H) are presented. The rotation-vibration-electronic (rovibronic) line lists, named \texttt{XAB}, consider transitions involving the \X, \A, and \BBp\ electronic states in the 0--30\,000~cm$^{-1}$ region (wavelengths $\lambda > 0.33$~$\mu$m) and are suitable for temperatures up to 5000 K. A comprehensive analysis of the published spectroscopic literature on CaH and MgH is used to obtain new extensive datasets of accurate rovibronic energy levels with measurement uncertainties and consistent quantum number labelling. These datasets are used to produce new spectroscopic models for CaH and MgH, composed of newly empirically-refined potential energy curves and couplings in/between the different electronic states (e.g.\ spin-orbit, electronic angular momentum, Born-Oppenheimer breakdown, spin-rotation, $\Lambda$-doubling) and previously published \textit{ab initio} transition dipole moment curves. Along with Einstein $A$ coefficients, state lifetimes and Land\'e $g$-factors are provided, the latter being particularly useful as CaH and MgH can be used to probe stellar magnetic fields. Computed energy levels have been replaced with the more accurate empirical values (if available) when post-processing the line lists, thus tailoring the line lists to high resolution applications. The \texttt{XAB} line lists are available from the ExoMol database at \href{http://www.exomol.com}{www.exomol.com} and the CDS astronomical database.
\end{abstract}


\section{Introduction}

Calcium monohydride (CaH) and magnesium monohydride (MgH) are important astrophysical molecules of particular relevance to brown dwarf, late-type stellar and planetary atmospheres. Since their original detection in sunspots over a century ago~\citep{08Olmste.CaH,07Fowler.MgH,09Fowler.MgH}, their spectra have been observed in a diverse range of environments, often providing considerable insight into the surrounding conditions in which they were found. For example, CaH and MgH are regularly used in the spectral classification of dwarf stars at higher temperatures (above $\approx 1500$~K)~\citep{05Kixxxx.dwarfs} and can help probe stellar magnetic fields on active G-K-M dwarfs~\citep{15AfBexx}. MgH features have been used to determine the surface gravity of the red giant star Arcturus~\citep{85BeEdGu.MgH} and several other cool giant stars~\citep{93BoBexx.MgH}. There is also speculation that CaH and MgH should occur in molecular clouds where they are thought to be a source of interstellar Mg and Ca, although searches have so far been unsuccessful~\citep{98SaWhKa.CaH}.

The detection of CaH or MgH in the context of exoplanets is expected and they are often considered in searches with other metal oxides and hydrides~\citep{17SeBoMa.TiO}. Metal-rich species like MgH can potentially cause strong thermal inversions in hot Jupiters if present in significant enough quantities~\citep{Gandhi:2019}. Central to any future detection in exoplanets is the underlying molecular line lists, which are usually converted to pressure- and temperature-dependent cross-sections~\citep{jt782,jt801,jt819} to serve as input opacities to atmospheric radiative transfer models. As was recently shown, cross-sections derived from different CaH and MgH line lists can produce noticeable differences in the atmospheric observables of hot Jupiters and M-dwarfs~\citep{21GhIyLi}. It is therefore essential that the underlying molecular line lists are of the highest possible quality to ensure the success of upcoming missions aimed at detecting and characterizing exoplanets. There are a number of line lists available for CaH and MgH which we will now discuss. Note that the main isotopologues are $^{40}$Ca$^{1}$H and $^{24}$Mg$^{1}$H, while $^{25}$Mg$^{1}$H and $^{26}$Mg$^{1}$H are stable minor isotopologues.

Rotation-vibration line lists in the \X\ ground electronic state of CaH and MgH (including the isotopologues $^{25}$MgH and $^{26}$MgH) covering microwave and infrared (IR) wavelengths were previously generated by \citet{jt529} for the ExoMol database~\citep{jt810}. The line lists are applicable for temperatures up to 2000~K and have been used in a number of studies of exoplanetary atmosphees~\citep{17SeBoMa.TiO,Juncher:2017,Malik:2019,Gandhi:2019} and stellar spectra~\citep{Hawkins:2016,Conroy:2018}. A natural extension to the work of \citet{jt529}, which is undertaken here, is to consider rotation-vibration-electronic (rovibronic) transitions from the \X\ ground state to the low-lying electronic states, notably the \A\ and \B\ states of CaH, and the \A\ and \Bp\ of MgH. The $A$--$X$ band of MgH is particularly important for establishing the isotopic abundances of Mg in stars~\citep{80ToLaxx.MgH,88McLaxx.MgH,00GaLaxx.MgH,03DaYDaL.MgH}, while $B^{\prime}$--$X$ lines can also be useful in this respect when the $A$--$X$ transitions are too strong to give meaningful results~\citep{99WaHiLi.MgH}.

The most recent rovibronic line list of CaH is that of \citet{18AlShxx.CaH}, which was generated using newly determined spectroscopic constants for the line positions and \textit{ab initio} transition dipole moment curves (TDMCs)~\citep{17ShAlRa.CaH} for the Einstein $A$ coefficients. A total of 11\,069 rovibronic transitions for CaH (and 14\,942 transitions for CaD) with total angular momentum quantum number $J \leq 53.5$ were computed between the $v=0$--4 vibrational levels of the \X\ electronic state to the $v=0$--3 levels of the \A\ state and $v=0$--2 levels of the \B\ state. This covered transitions with wavenumbers between 9480--$18\,460$~cm$^{-1}$ (wavelengths $1.06> \lambda > 0.54$~$\mu$m).

Another commonly used CaH rovibronic line list is that of \citet{03WeStKi.CaH}, which was computed using the \textit{ab initio} potential energy curves (PECs) and TDMCs of \citet{95LeJexx.CaH}. Transition energies and oscillator strengths are available up to 35\,000~cm$^{-1}$ (wavelengths $\lambda > 0.29$~$\mu$m) from the \X\ ground state to the low-lying \A, \B, \C, \D\ and \E\ electronic states. Although boasting good coverage, the calculated line positions are of limited accuracy given that the PECs were not rigorously refined to experimental data with only minor empirical adjustments to the dissociation energies of the curves. Furthermore, fine structure terms that lead to the splitting of spectral lines such as spin-orbit or spin-rotation coupling were neglected in their theoretical model.

For the main isotopologue of MgH, the $^{24}$MgH line list of \citet{13GhShBe.MgH} is one of the most accurate to date and contains around 31\,000 rovibronic transitions from the $v=0$--11 vibrational levels of the \X\ electronic state to the $v=0$--3 levels of the \A\ state and $v=0$--9 of the \Bp\ state. A complete list of line positions up to $J=50.5$ covering transitions between 8900--$28\,480$~cm$^{-1}$ (wavelengths $1.12> \lambda > 0.35$~$\mu$m) was derived from experimentally determined term values~\citep{07ShHeLe.MgH}, while Einstein $A$ coefficients were computed using experimentally determined PECs~\citep{11ShBexx.MgH} and \textit{ab initio} TDMCs~\citep{12MoShxx.MgH}. Further spectroscopic measurements of the less common isotopologues of MgH by \citet{13HiWaRa.MgH} have led to improved $A$--$X$ line lists for $^{25}$MgH and $^{26}$MgH that helped identify new transitions in sunspots and metal-poor dwarf and giant stars. A multi-isotopologue, direct-potential-fit analysis of $A$--$X$ and $B^{\prime}$--$X$ emission spectra has also been performed to produce a highly accurate \X\ ground state PEC~\citep{13HeShTa.MgH} of MgH.

Similar to CaH, Weck and co-workers have generated theoretical line lists for the $A$--$X$~\citep{03WeScSt.MgHline} and $B^{\prime}$--$X$~\citep{03SkWeSt.MgH} bands of MgH but again these line lists are of limited accuracy having been generated with \textit{ab initio} PECs with only slight adjustment to experimental data. Both the CaH and MgH line lists by Weck and co-workers are available from the University of Georgia Molecular Opacity Project (UGAMOP) (see the website \href{https://www.physast.uga.edu/ugamop/}{www.physast.uga.edu/ugamop/}; accessed November 2021). Finally, it is worth noting that the MoLLIST empirical line list database~\citep{MOLLIST} of Peter Bernath, who was involved in many of the spectroscopic measurements of CaH and MgH discussed above, hosts data on CaH and MgH. These line lists, along with many other diatomic species, have been conveniently formatted for inclusion in the ExoMol database and further details can be found in \citet{jt790}. 

Both the MoLLIST and \citet{jt529} line lists of CaH and MgH have been used in the recent EXOPLINES molecular absorption cross-section database for brown dwarf and giant exoplanet atmospheres~\citep{21GhIyLi}. Combining different line lists to gain proper wavelength coverage is common practice in the atmospheric modelling of exoplanets. More desirable, however, would be to have a single, comprehensive molecular line list that is accurate and complete over an extended wavelength region. The property of completeness of a line list is particularly important for characterizing exoplanet atmospheres and correctly modelling molecular opacities~\citep{jt572}. Motivated by recent developments in the methodology of line list construction for the ExoMol database~\citep{jt626}, we thus find it worthwhile to produce new rovibronic line lists for CaH and MgH that extend to visible and near-ultraviolet wavelengths and that have been adapted for the high-resolution spectroscopy of exoplanets~\citep{14Snellen,18Birkby}. The adaptation of molecular line lists for high-resolution applications is a major activity for improving the ExoMol database, see \citet{jt835} for example.

It is worth mentioning that the alkaline-earth monohydrides are promising candidates for laser-cooling~\citep{04DiRosa.CaH,14GaGaxx.MgH,20Gaoxxx.MgH,21YiYaYi.CaH} and a detailed knowledge of their rovibronic energy level structure and transition properties, especially concerning the \A--\X\ band, could assist this field when designing efficient laser-cooling schemes.

\section{Methods}
\label{sec:methods}

\subsection{MARVEL analysis}

The MARVEL (Measured Active Rotational-Vibrational Energy Levels) procedure~\citep{jt412,07CsCzFu.method,12FuCsxx.methods,jt750} has become an indispensable part of the line list construction process. The program MARVEL (available via the online app at \href{http://kkrk.chem.elte.hu/marvelonline}{http://kkrk.chem.elte.hu/marvelonline}; accessed November 2021) takes as input a user-constructed dataset of assigned spectroscopic transitions with measurement uncertainties and converts them into a consistent set of labelled empirical-quality energy levels with the uncertainties propagated from the input transitions to the output energies. The resulting energy levels have two main applications. Firstly, they are used to refine the PECs and state coupling curves of the theoretical spectroscopic model of the molecule in line list calculations. Secondly, when post-processing the computed line list the calculated energy levels can be substituted with the equivalent MARVEL values if available, further improving the accuracy of the predicted line positions.

For CaH, 3663 experimental wavenumbers were taken from \citet{13ShRaBe.CaH}, which covered transitions between the $v=0$--4 vibrational levels of the \X\ electronic state to the $v=0$--3 levels of the \A\ state and $v=0$--2 levels of the \B\ state. The data from \citet{13ShRaBe.CaH} included $B$--$X$ lines from \citet{76BeKlMa.CaH} and ground state data from \citet{04ShWaGo.CaH}. Transitions involving highly excited vibrational levels of the \B\ electronic state are available~\citep{16WaYoUc.CaH,18WaTaKo.CaH}, however, their inclusion into the input CaH MARVEL dataset led to inconsistencies with the data from \citet{13ShRaBe.CaH}, and they were therefore discarded. The input MARVEL dataset was processed using the Cholesky (analytic) approach with a 0.05~cm$^{-1}$ threshold on the uncertainty of the ``very bad'' lines,  producing 1260 energy levels up to $N \leq 55$ below 25\,323~cm$^{-1}$, where $N$ is the rotational angular momentum quantum number. Five quantum numbers were used to uniquely assign the CaH MARVEL energy levels: an electronic state label ($X$, $A$, $B$), $N$, the vibrational state $v$, the rotationless parity $e$/$f$, and an id number ranging from 1--13 used in \citet{13ShRaBe.CaH} (see the Table S1 description in supplementary material of \citet{13ShRaBe.CaH} for definitions), which is related to $v$ and the quantum labels $F_1$ and $F_2$, which denote spin components $J = N + 1/2$ and $J = N - 1/2$, respectively, where $J$ is the total angular momentum quantum number.

MARVEL datasets were produced for the three isotopologues of MgH with the majority of transition data coming from \citet{13HeShTa.MgH}, a study which performed a multi-isotopologue fit of experimental transition data to determine an accurate \X\ ground state PEC of MgH. In \citet{13HeShTa.MgH}, new $B^{\prime}$--$X$ measurements for $^{25}$MgH and $^{26}$MgH were performed and analysed alongside previous $A$--$X$ and $B^{\prime}$--$X$ measurements of $^{24}$MgH~\citep{07ShHeLe.MgH}, and a range of $X$--$X$ ground state data~\citep{04ShApGo.MgH,86LeZiEv.MgH,88LeDeDe.MgH,90ZiJeEv.MgH,93ZiBaAn.MgH,07ShHeLe.MgH}. This served as a valuable source of labelled transition data on all three isotopologues that was already formatted with a consistent set of quantum numbers.

For $^{24}$MgH, we extracted 7453 transitions from \citet{13HeShTa.MgH} covering the $v=0$--11 (\X) and $v=0$--3 (\A, \Bp) levels up to $N \leq 49$. In addition, 29 low-$J$ transitions of the \A($v=0$)--\X($v=0$) band were taken from \citet{14ZhStxx.MgH}, along with 101 $B^{\prime}$--$X$ transitions  from \citet{11ShBexx.MgH} involving the $v=4$ (\Bp) level up to $N \leq 21$, and 937 $B^{\prime}$--$X$ transitions involving the $v=5$--9 (\Bp) levels up to $N \leq 31$ from \citet{78BaLixx.MgH}. The final $^{24}$MgH MARVEL dataset contained 8520 experimental transition wavenumbers resulting in 1856 empirical-quality energy levels up to $N \leq 49$ below 29\,748~cm$^{-1}$. Four quantum numbers were used to uniquely identify the energy levels of MgH: an electronic state label ($X$, $A_1$, $A_2$, $B$) where $A_1$ and $A_2$ correspond to the $^2\Pi_{1/2}$ and $^2\Pi_{3/2}$ components of the \A\ state caused by spin-orbit splitting, $N$, $v$, and $e$/$f$.

For $^{25}$MgH ($^{26}$MgH), 1046 (913) transitions were extracted from \citet{13HeShTa.MgH} covering the $v=0$--8 (\X) and $v=0$--1 (\Bp) levels up to $N \leq 36$, along with 311 (324) transitions from \citet{13HiWaRa.MgH} involving the $v=0,1$ (\A) levels up to $N \leq 35$. The MARVEL datasets produced 729 energy levels for each isotopologue. All MARVEL input transitions and output energy files are given as part of the supplementary material.

\section{Spectroscopic Models}

\subsection{CaH}

\subsubsection{Potential energy and coupling curves}

The PECs and various couplings between the electronic states were established through rigorous empirical refinement to the CaH MARVEL dataset of rovibronic energy levels using the computer program \Duo~\citep{jt609}. Full details of the analytic representations used for the PECs and different couplings are given in the supplementary material and only a brief summary is provided here. The \Duo\ input files for CaH and MgH, which contains all the parameters and defines the spectroscopic model, are provided in the supplementary material and can be found on the ExoMol website. The \Duo\ online manual (see \href{https://duo.readthedocs.io/en/latest/index.html}{https://duo.readthedocs.io/en/latest/index.html}; accessed November 2021), is another valuable resource for details of the relevant keywords, parameters and methodologies.

\begin{figure}
\centering
\includegraphics[width=0.5\textwidth]{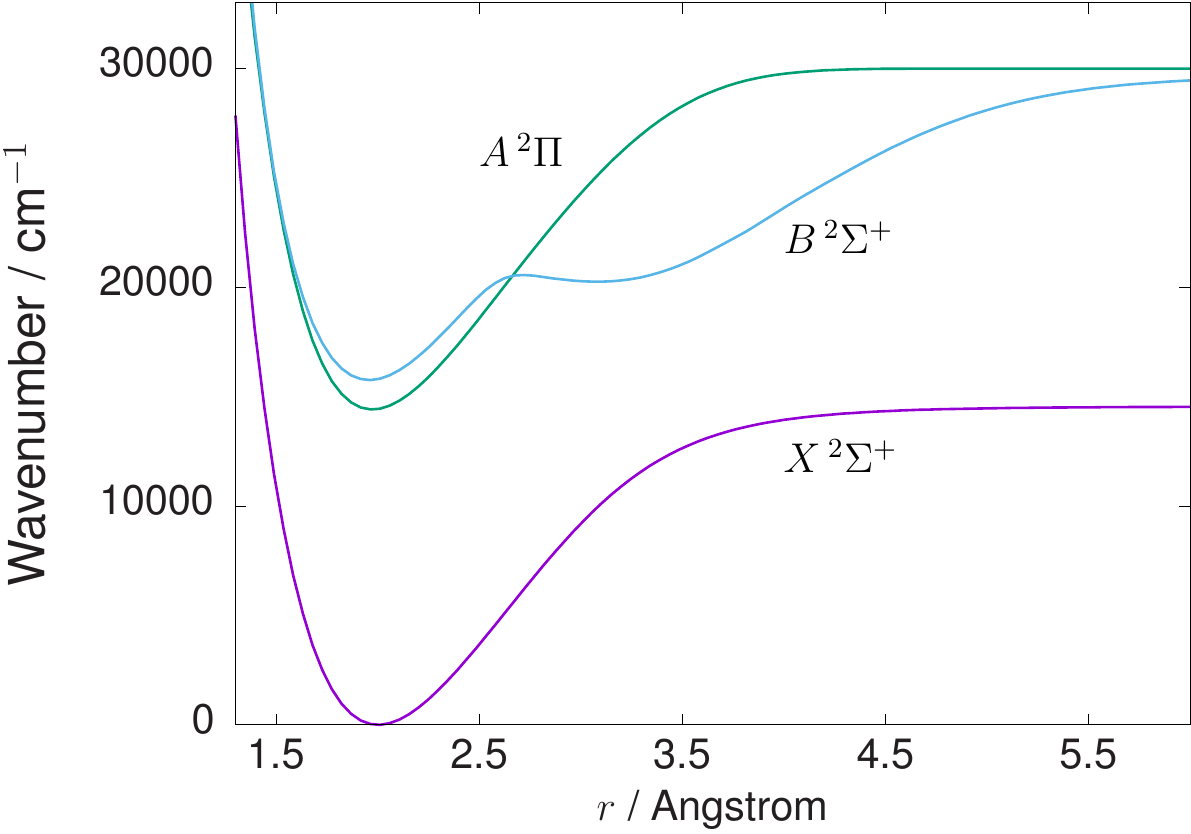}
\caption{\label{fig:pec_cah}The (adiabatic) potential energy curves of the \X, \A\ and \B\ electronic states of CaH.}
\end{figure}

The \X\ ground state of CaH was represented as a Morse/Long-Range (MLR) potential function ~\citep{MLRpaper} with the parameter values from \citet{jt529} used as the starting point in the refinement, while the \A\ and \B\ PECs were represented as Extended Morse-Oscillator (EMO) functions~\citep{EMO}. The (adiabatic) PECs of CaH are illustrated in Fig.~\ref{fig:pec_cah}. Accurately modelling the \B\ state of CaH is  challenging due to an avoided crossing that occurs around the Ca--H bond length value of $r_{\mathrm{CaH}}=2.5$~\AA, causing the kink and second well in the curve. To correctly model this behaviour it was necessary to include a ``dummy'' \D\ state with an associated $D$--$A$ spin-orbit and electronic angular momentum coupling and $D$--$B$ diabatic coupling term. Although the \D\ of CaH is physical, in this work it should not be considered reliable and no transitions to it were calculated.

Different couplings between the electronic states are usually needed in the spectroscopic model to achieve the highest possible accuracy for computed energy levels. The following were included for CaH: $A$--$A$, $B$--$A$ and $D$--$A$ spin-orbit coupling, $B$--$A$ and $D$--$A$ electronic angular momentum coupling, Born-Oppenheimer breakdown functions and spin-rotational coupling in the \X, \A\ and \B\ states, and $\Lambda$-doubling coupling in the \A\ state.

For the PECs and most of the couplings, the parameters were determined by directly fitting to the experimental energy levels in a weighted least-squares fitting procedure. Energies with smaller uncertainties were given a higher weighting in the fit while energy levels that had only been involved in one transition (information obtained in the MARVEL procedure), and therefore cannot be classed as reliable, were assigned with lower weights. Fitting of the $A$--$A$, $B$--$A$ spin-orbit and $B$--$A$ electronic angular momentum coupling curves was slightly different in that these were based on calculated \textit{ab initio} curves that were subsequently ``morphed'' (essentially shifted by a constant factor or simple function) to agree with experiment. \textit{Ab initio} calculations were performed using state-averaged multi-configurational self-consistent field (MCSCF) theory~\citep{85KnWexx.ai,85WeKnxx.ai} involving the \X, \A\ (both $^2\Pi_{1/2}$ and $^2\Pi_{3/2}$ spin components) and \B\ states in conjunction with the correlation consistent basis sets cc-pCVQZ for Ca~\citep{02KoPexx.CaOH} and cc-pVQZ for H~\citep{89Dunning.ai}. The active space included 11 electrons distributed between $(5\,a_1,2\,b_1,2\,b_2,0\,a_2)$ orbitals in $\bm{C}_{\mathrm{2v}}$ point group symmetry. Calculations were done with the quantum chemistry program MOLPRO2015~\citep{MOLPRO,Molpro:JCP:2020} on a grid of Ca--H bond length $r_{\mathrm{CaH}}=1.6$--5.0~\AA.

\begin{figure}
\centering
\includegraphics[width=0.5\textwidth]{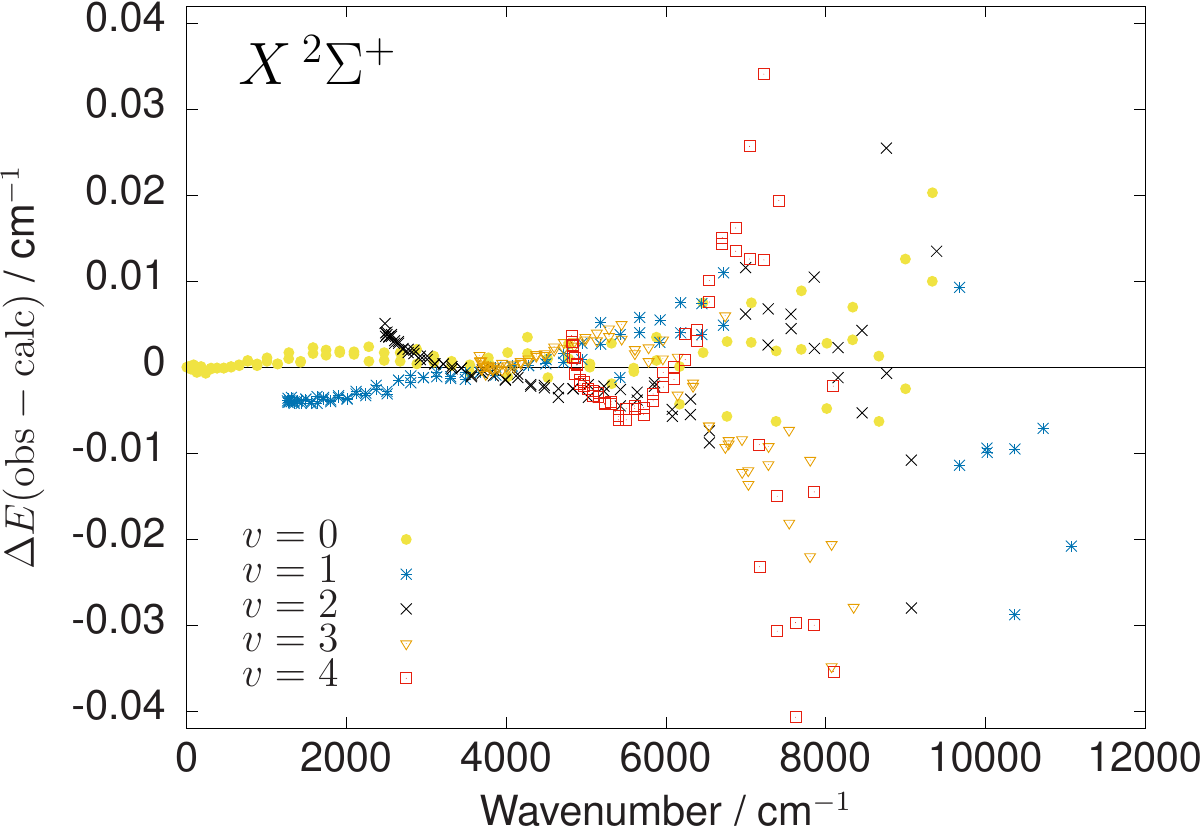}
\includegraphics[width=0.5\textwidth]{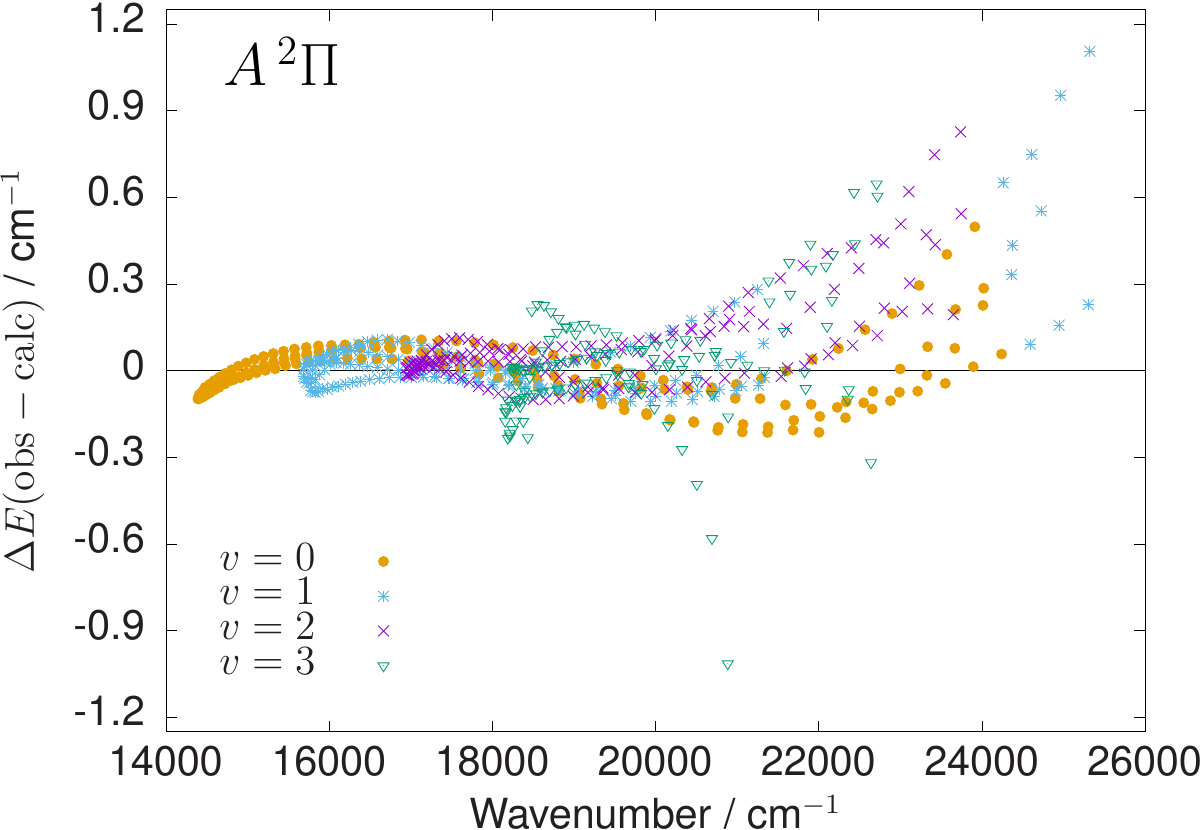}
\includegraphics[width=0.5\textwidth]{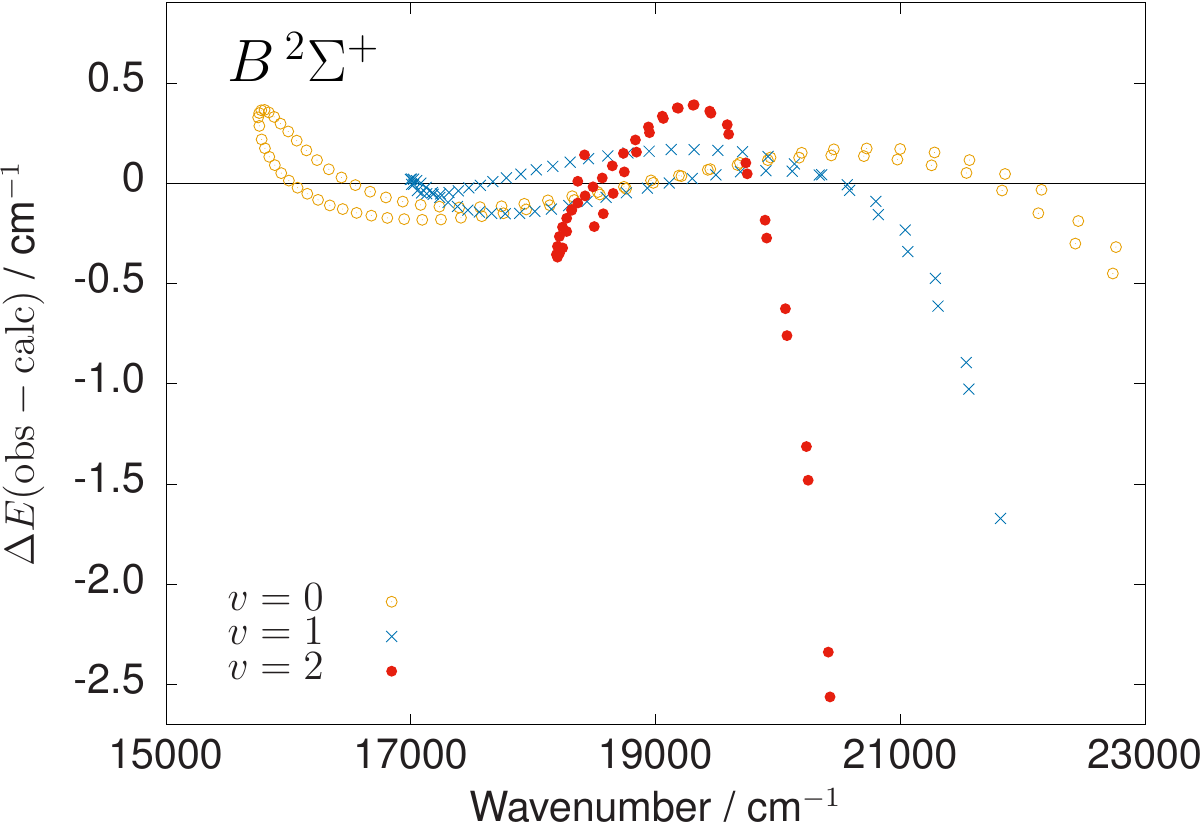}
\caption{\label{fig:res_pec_cah}Residual errors $\Delta E(\mathrm{obs}-\mathrm{calc})$ between the empirically-derived MARVEL energy levels of $^{40}$CaH (\X, \A, and \B) and the calculated \textsc{Duo} values using the refined spectroscopic model for the different vibrational levels $v$.}
\end{figure}

All curves were adjusted in the refinement, reproducing 1170 term values up to $J=54.5$ with a weighted root mean square (w-rms) error of 0.001~cm$^{-1}$ and root mean square (rms) error of 0.241~cm$^{-1}$. The results of the refinement are illustrated in Fig.~\ref{fig:res_pec_cah}. The residual errors $\Delta E(\mathrm{obs}-\mathrm{calc})$ (in cm$^{-1}$) between the empirically-derived MARVEL energy levels and the calculated \textsc{Duo} values from the refined spectroscopic model show the expected behaviour. In each electronic state, the errors are larger for higher vibrational and rotational excitation (corresponding to higher energies). These states are usually not as well characterised in experiment and therefore have lower weights and less importance in the refinement.

\subsubsection{Dipole moment curves}

For the $A$--$X$ and $B$--$X$ bands, we have used the TDMCs of \citet{17ShAlRa.CaH}, which were generated at a high-level of \textit{ab initio} theory (multireference configuration interaction with a quadruple-zeta quality correlation consistent basis set, MRCI/cc-pwCVQZ) on a large grid of Ca--H bond length $r_{\mathrm{CaH}}=2.0$--14.0~$a_0$. These dipoles have been utilised in recent rovibronic line list calculations~\citep{18AlShxx.CaH}. The ground state \X\ DMC from the previous ExoMol study of \citet{jt529} was also employed. This was computed at a high-level of \textit{ab initio} theory (coupled cluster with a quintuple-zeta quality correlation consistent basis set, RCCSD(T)/cc-pCV5Z) and guarantees similar intensity predictions to the original ExoMol CaH line list in the microwave and IR. 

\subsubsection{\Duo\ calculations}

Rovibronic line list calculations were carried out with the computer program \Duo, which variationally solves the diatomic molecular Schr\"odinger equation. \Duo\ has been extensively used by the ExoMol project and there is a range of literature available on its methodologies~\citep{jt632,jt609,jt626} and its application to other diatomic molecules (see previous ExoMol line list publications). Here, we only summarise the key calculation parameters.

A grid-based sinc discrete variable representation (DVR) method employing 501 grid points uniformly distributed in the range $r_{\mathrm{CaH}}=1.3$--6.0~\AA\ was utilised to solve the coupled Schr\"odinger equation. The basis set contained vibrational levels up to $v_{\mathrm{max}}=15,20,25$ for the $X,A,B$ electronic states, respectively, which ensures converged energies below the dissociation limit of each electronic state. Transitions were computed with a lower state energy threshold of $h c \cdot 13\,700$~cm$^{-1}$ ($h$ is the Planck constant and $c$ is the speed of light), which roughly corresponds to the  dissociation energy ($D_0$) of the \X\ state where we have calculated the zero-point energy (ZPE) to be 644.6~cm$^{-1}$. An upper state energy threshold of $h c \cdot 29\,900$~cm$^{-1}$ was selected so as to be just below the dissociation asymptote of the \A\ and \B\ states. 

A total of 293\,151 transitions up to $J=61.5$ were computed between 6825 energy levels for the CaH line list. Calculations used atomic mass values 39.962590863000~Da ($^{40}$Ca) and 1.00782503223~Da ($^{1}$H).

\subsection{MgH}

\subsubsection{Potential energy and coupling curves}

\begin{figure}
\centering
\includegraphics[width=0.5\textwidth]{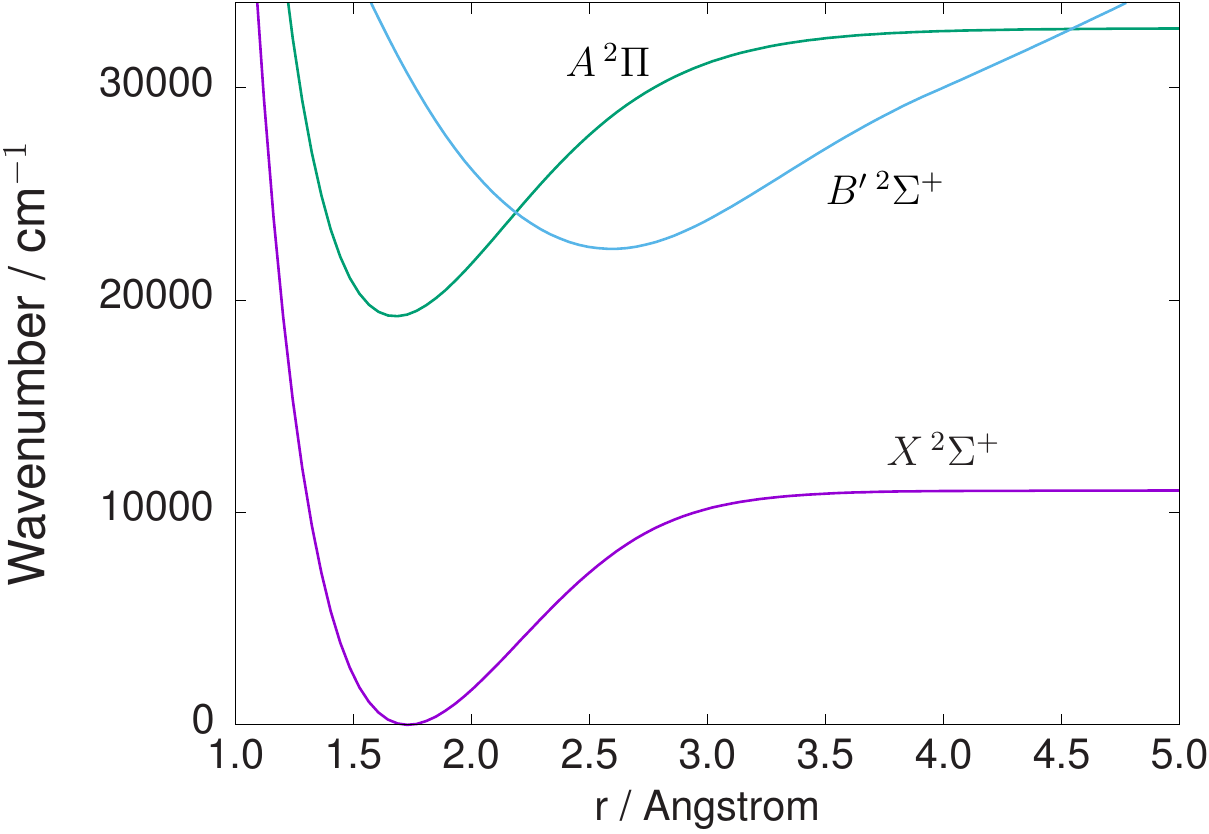}
\caption{\label{fig:pec_mgh}The empirically refined (adiabatic) potential energy curves of MgH. Note that the \Bp\ state displays incorrect dissociation behaviour (see text).}
\end{figure}

We have adopted the highly accurate \X\ ground state MLR potential function of \citet{13HeShTa.MgH} and performed a minor empirical refinement of the parameters to obtain better agreement with higher $J$ rotational states. The \X\ state also required Born-Oppenheimer breakdown and spin-rotational coupling terms to further improve its description. An EMO analytic function was used for the \A\ excited state and the parameters were established by empirical refinement. Accurately describing the \Bp\ PEC was more challenging due to a shallower minimum that is located at a different equilibrium bond length compared to the \X\ and \A\ states. For this reason, the \Bp\ state was modelled using an empirically determined Rydberg–Klein–Rees (RKR) PEC~\citep{11ShBexx.MgH}, which was subsequently morphed in the refinement process. The refined PECs of MgH are illustrated in Fig.~\ref{fig:pec_mgh}. As a result of the morphing procedure and the fact that the RKR \Bp\ PEC is only defined up to an Mg--H bond length value of $r_{\mathrm{MgH}}\approx 4.0$~\AA, the \Bp\ state PEC exhibits incorrect dissociation behaviour and should actually follow the dissociation limit of the \A\ state. However, as this region of the \Bp\ PEC lies above 30\,000~cm$^{-1}$ which is the upper state energy threshold of our line list calculations, we can safely ignore this incorrect dissociation behaviour.

The $A$--$X$, $A$--$A$ and $B^{\prime}$--$A$ spin-orbit and $A$--$X$ and $B^{\prime}$--$A$ electronic angular momentum coupling curves were calculated \textit{ab initio} and then morphed in the refinement. The couplings were computed using MOLPRO2015~\citep{MOLPRO,Molpro:JCP:2020} on a grid of $r_{\mathrm{MgH}}=1.1$--5.0~\AA\ using state-averaged MCSCF theory over the \X, \A\ (both $^2\Pi_{1/2}$ and $^2\Pi_{3/2}$ spin components) and \Bp\ states in conjunction with the correlation consistent basis sets cc-pCVQZ for Mg~\citep{11PaWoPe.ai} and cc-pVQZ for H~\citep{89Dunning.ai}. The active space contained 3 electrons distributed between $(5\,a_1,2\,b_1,2\,b_2,0\,a_2)$ orbitals in $\bm{C}_{\mathrm{2v}}$ point group symmetry.

\begin{figure}
\centering
\includegraphics[width=0.5\textwidth]{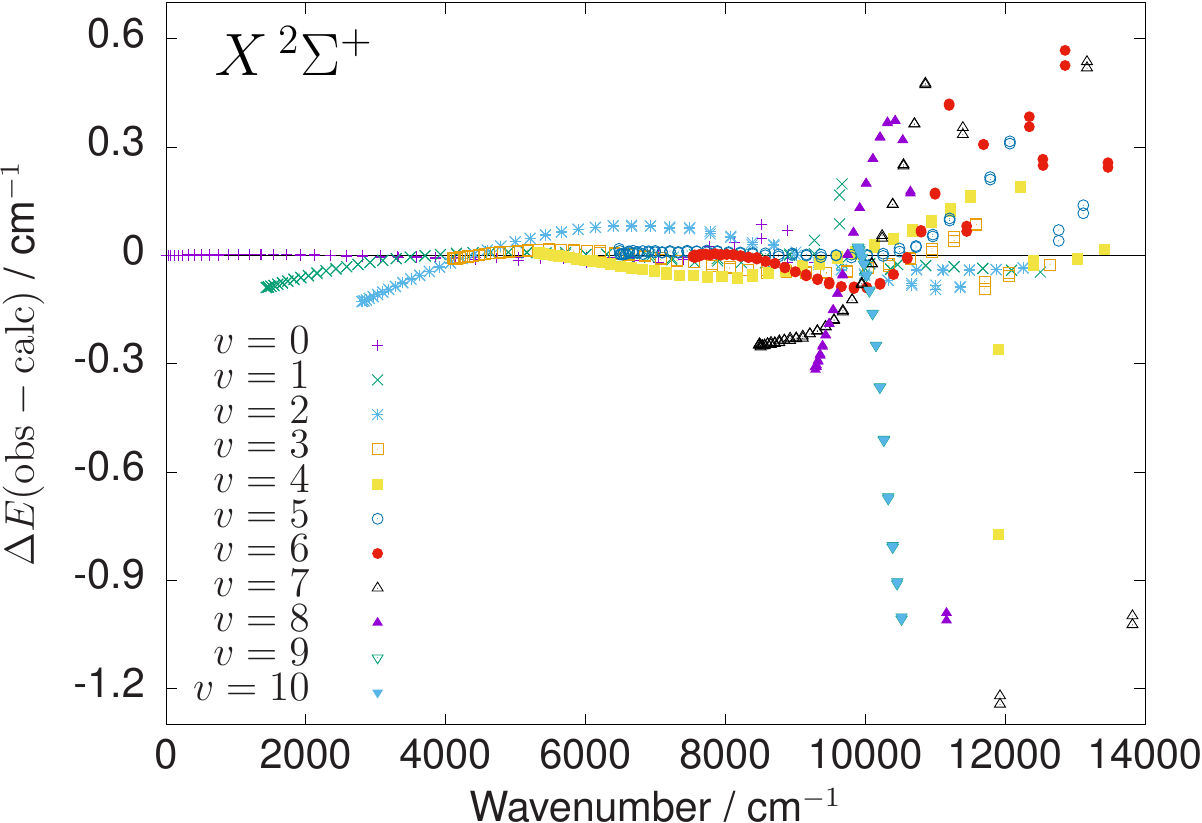}
\includegraphics[width=0.5\textwidth]{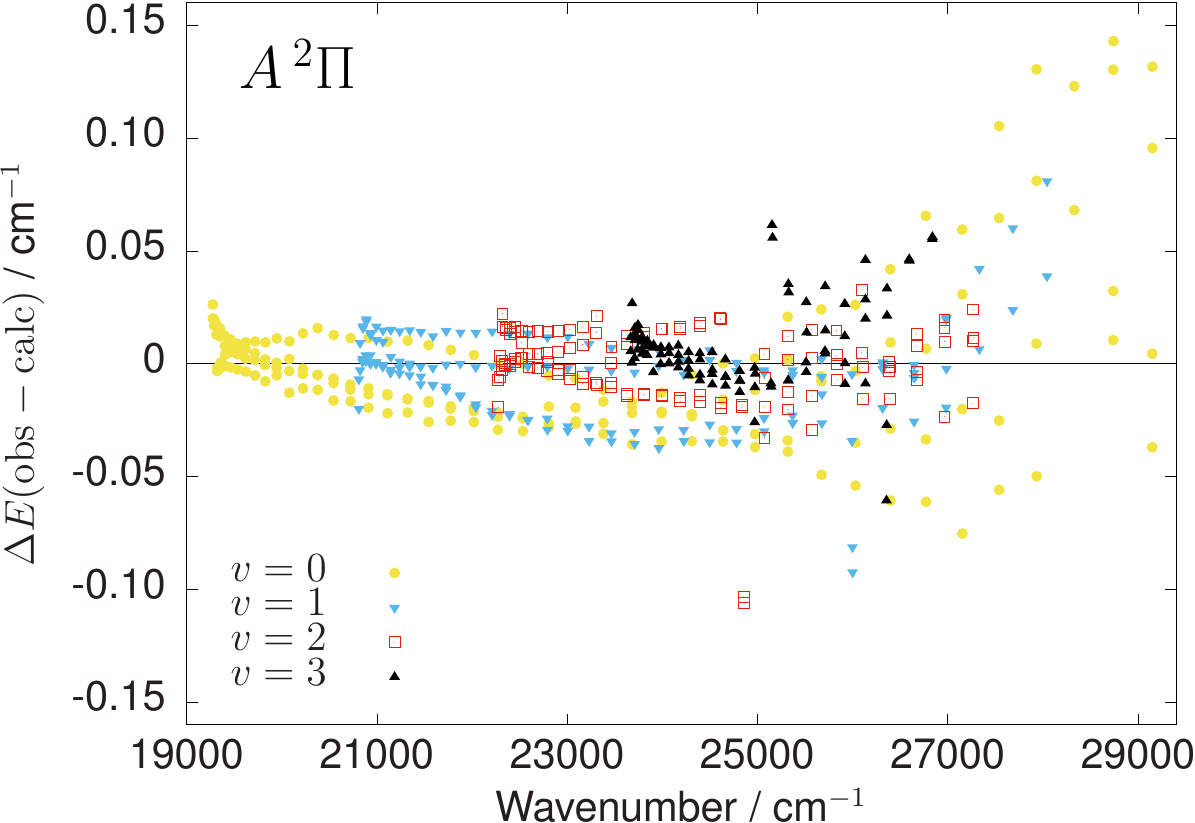}
\includegraphics[width=0.5\textwidth]{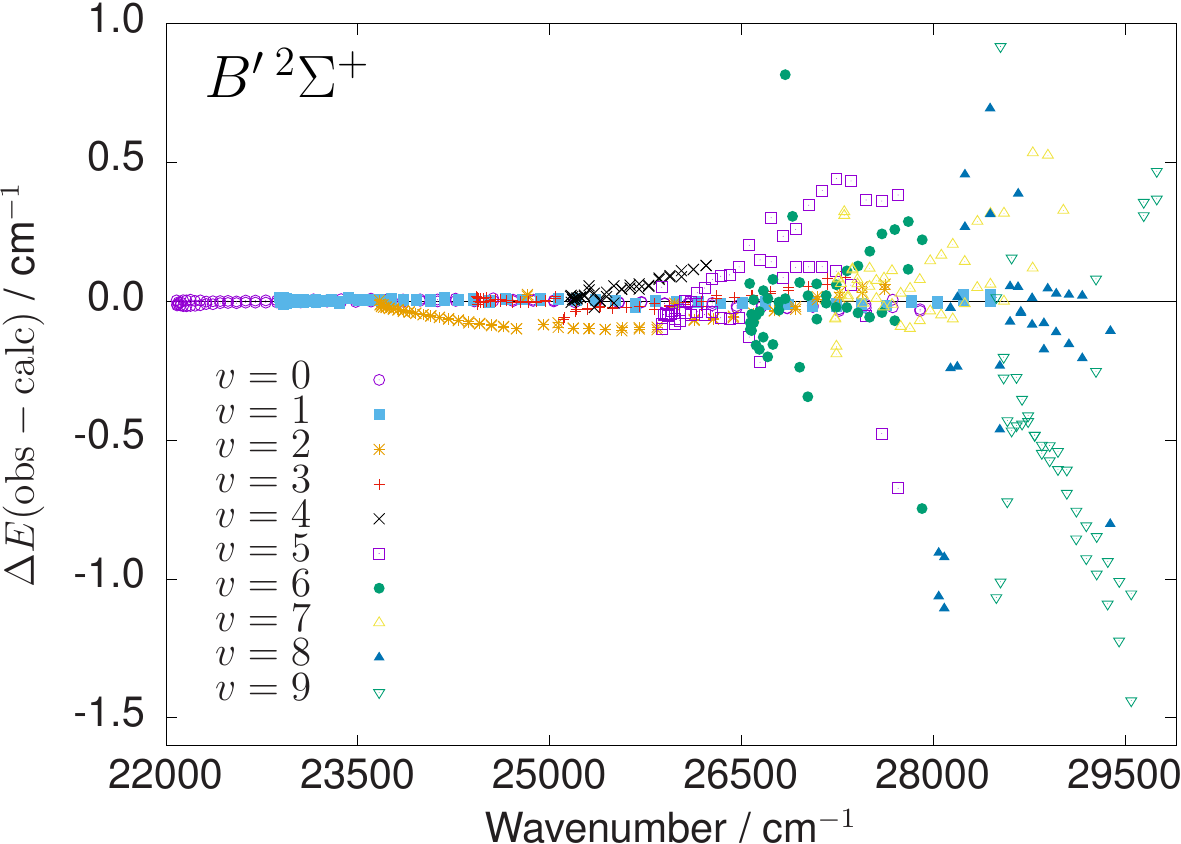}
\caption{\label{fig:res_pec_mgh}Residual errors $\Delta E(\mathrm{obs}-\mathrm{calc})$ between the empirically-derived MARVEL energy levels of $^{24}$MgH (\X, \A, and \Bp) and the calculated \textsc{Duo} values from the refined spectroscopic model.}
\end{figure}

For $^{24}$MgH, all curves and couplings were refined to the MARVEL dataset of energy levels. A total of 1827 energy levels up to $J=49.5$ were weighted in the refinement of $^{24}$MgH and were reproduced with a w-rms error of 0.019~cm$^{-1}$ and rms error of 0.262~cm$^{-1}$. The results of the refinement of $^{24}$MgH are illustrated in Fig.~\ref{fig:res_pec_mgh}, where we have plotted the residual errors $\Delta E(\mathrm{obs}-\mathrm{calc})$ (in cm$^{-1}$) between the MARVEL energies and the final \Duo\ calculated values using the refined spectroscopic model. Like CaH, the residual errors get larger with increasing energy in each electronic state, usually as these levels correspond to higher $J$ states that are less well characterised and thus have lower weights and less importance in the refinement. The fitting errors for excited vibrational states tend to be larger than those of lower vibrational states, again for similar reasons.

The spectroscopic models of the isotopologues $^{25}$MgH and $^{26}$MgH were constructed independently as they each possessed their own MARVEL dataset of energy levels. An additional set of empirically-derived term values~\citep{13HeShTa.MgH,13HiWaRa.MgH} were used to ``plug'' gaps in the isotopologue MARVEL datasets. These additional energies were calculated using an effective Hamiltonian model analysing the same transition data used in our MARVEL analysis but as these energy levels were extrapolated and not ``observed'', they were given lower weights in the refinement process.

Using the $^{24}$MgH spectroscopic model as a starting point for the isotopologues, it was only necessary to adjust coupling terms dependent on the nuclear masses instead of refining all of the PECs and couplings. Only 12 parameters were allowed to vary, namely those associated with the $A$--$X$ and $A$-$A$ electronic angular momentum coupling, $X$--$X$ and $A$--$A$ Born-Oppenheimer breakdown terms, and $X$--$X$ spin-rotation coupling. Similar levels of accuracy were obtained for the isotopologue refinements. For $^{25}$MgH, 1073 energy levels up to $J=39.5$ were reproduced with a w-rms error of 0.021~cm$^{-1}$ and rms error of 0.743~cm$^{-1}$, while 1083 energy levels up to $J=39.5$ were reproduced with a w-rms error of 0.021~cm$^{-1}$ and rms error of 0.799~cm$^{-1}$ for $^{26}$MgH.

\subsubsection{Dipole moment curves}

We have employed the TDMCs of \citet{12MoShxx.MgH} for the $A$--$X$ and $B^{\prime}$--$X$ bands. These were computed using a high-level of \textit{ab initio} theory (multireference configuration interaction with a quadruple-zeta quality correlation consistent basis set, MRCI/aug-cc-pCVQZ) on a large grid of Mg--H bond length $r_{\mathrm{MgH}}=2.2$--20.0~$a_0$ and were previously used in $^{24}$MgH line list calculations~\citep{13GhShBe.MgH}. Similar to the CaH calculations, we have taken the \X\ ground state DMC from the previous ExoMol study~\citep{jt529} to ensure the same intensity predictions as the original ExoMol MgH line list in the microwave and IR. Similarly, this DMC was computed at a high-level of \textit{ab initio} theory (coupled cluster with a quintuple-zeta quality correlation consistent basis set, RCCSD(T)/cc-pCV5Z).

\subsubsection{\Duo\ calculations}

The MgH \Duo\ calculations were performed on a grid of 401 points in the range $r_{\mathrm{MgH}}=1.0$--5.0~\AA\ using a basis set containing vibrational levels up to $v_{\mathrm{max}}=25,30,30$ for the $X,A,B$ electronic states, respectively. The use of a large basis set provided the best achievable accuracy in the refinement and although this included states above the dissociation limit, these were removed when post-processing the MgH line lists (see text below). Calculations employed atomic mass values of 23.985041697~Da ($^{24}$Mg), 24.985836976~Da ($^{25}$Mg), 25.982592968~Da ($^{26}$Mg) and 1.00782503223~Da ($^{1}$H).

The highest bound vibrational level of the \X\ ground state in MgH is the $v=11$ level~\citep{07ShHeLe.MgH}, which lies just below the dissociation asymptote. In \citet{07ShHeLe.MgH}, a number of ``quasibound'' states were observed in their measured MgH spectra at $T\approx 1500$~K and were characterised. These quasibound states exist above the zero-point dissociation energy ($D_{0}=10\,365 \pm 0.5$~cm$^{-1}$ for $^{24}$MgH~\citep{07ShHeLe.MgH}) but below the centrifugal barrier maximum, and they were present in our MARVEL datasets and utilised in the refinements. Since the highest observed quasibound energy level of the \X\ state of the $^{24}$MgH MARVEL dataset is at $h c \cdot 13\,469.7$~cm$^{-1}$, we have used a lower state energy threshold of $h c \cdot 13\,500$~cm$^{-1}$ in our line list calculations. However, we have not considered any vibrational levels above $v=11$ in the \X\ state. Transitions were computed with an upper state energy threshold of $h c \cdot 30\,000$~cm$^{-1}$. 

Overall, 88\,575 transitions up to $J=59.5$ were computed between 3148 energy levels for the $^{24}$MgH line list, 88\,776 transitions up to $J=59.5$ between 3156 energy levels for the $^{25}$MgH line list, and 88\,891 transitions up to $J=60.5$ between 3160 energy levels for the $^{26}$MgH line list.

\section{Results}
\label{sec:results}

\subsection{Line list format}

As standard, the CaH and MgH \texttt{XAB} line lists are provided in the ExoMol data format~\citep{jt810}, illustrated in Tables~\ref{tab:trans} and \ref{tab:states}. The \texttt{.trans} file, see Table~\ref{tab:trans} for an example from the CaH line list (the structure is identical for MgH), contains all the computed transitions with upper and lower state ID labels, Einstein $A$ coefficients (in s$^{-1}$) and transition wavenumbers (in cm$^{-1}$). Table~\ref{tab:states} shows an example of the CaH \texttt{.states} file (the structure is identical for the MgH line lists), which contains all the computed rovibronic energy levels (in cm$^{-1}$), each labelled with a unique state ID counting number and quantum number labelling. Since CaH and MgH are known to be suitable molecular probes of stellar magnetic fields~\citep{15AfBexx}, we have also computed Land\'e $g$-factors which describe the behaviour of molecular states in the presence of a weak magnetic field as given by the Zeeman effect. These can be routinely calculated in \Duo~\citep{jt656} and are listed in column 7 of the \texttt{.states} file after the calculated state lifetimes. 

Where available, calculated \Duo\ energy levels and their uncertainties have been replaced with the more accurate empirically-derived MARVEL values and this information is indicated in the \texttt{.states} file through the labels ``{\tt Ca}'' for calculated and ``{\tt Ma}'' for MARVEL. As is now standard practice for the ExoMol database, we provide estimated uncertainties on all  energy levels (hence transition wavenumbers). MARVEL uncertainties are given where appropriate and  uncertainties for the calculated energies were estimated using the expression,
\begin{equation}
\label{e:unc}
{\rm unc} = a + b v + c J(J+1),
\end{equation}
where $v$ corresponds to the vibrational level, $J$ is the total angular momentum quantum number of the state, and the constants $a=0.5$, $b=0.5$ and $c=0.01$. These should be regarded as conservative estimates and in many instances we expect the energy levels to be more accurate than the uncertainties might suggest. However, we prefer a more cautious approach and a way to ensure the user is aware of the difference in reliability between calculated and ``MARVELised'' lines, particularly in regard to high-resolution applications. For reference, all \Duo\ calculated energies are provided in the final column of the \texttt{.states} file.

\begin{table}
\centering
\caption{\label{tab:trans}Extract from the \texttt{.trans} file of the CaH line list.}
\tt
\centering
\begin{tabular}{rrrr}
\toprule\toprule
\multicolumn{1}{c}{$f$}	&	\multicolumn{1}{c}{$i$}	& \multicolumn{1}{c}{$A_{fi}$}	&\multicolumn{1}{c}{$\tilde{\nu}_{fi}$} \\
\midrule
798 &  853 & 5.5465E+05 & 13203.662260\\
7149 & 6786 & 1.6977E+02 & 13203.666513\\
60 & 304 & 6.6048E-03 & 13203.722911\\
798 & 633 &  1.8429E+04 & 13203.727224\\
60 & 105 &  1.3280E-02 & 13203.736727\\
\bottomrule\bottomrule
\end{tabular} \\ \vspace{2mm}
\rm
\noindent
$f$: Upper  state counting number;\\
$i$:  Lower  state counting number; \\
$A_{fi}$:  Einstein-$A$ coefficient (in s$^{-1}$); \\
$\tilde{\nu}_{fi}$: Transition wavenumber (in cm$^{-1}$).\\
\end{table}

\begin{table*}
\centering
\caption{\label{tab:states}Extract from the \texttt{.states} file of the CaH line list.}
{\tt  \begin{tabular}{rrrrrrrcclrrrrcr}
\toprule\toprule
$i$ & $\tilde{E}$ (cm$^{-1}$) & $g_i$ & $J$ & unc &  $\tau$ & $g$& \multicolumn{2}{c}{Parity} 	& State	& $v$	&${\Lambda}$ &	${\Sigma}$ & $\Omega$ & Label & Calc. \\
\midrule
1 & 0.000000 & 4 &     0.5 &     0.000001 &         Inf &   2.002313 & + &  e & X2Sigma+ &     0 &   0 &  0.5 &  0.5 & Ma &    0.000000\\
2 & 1260.127299 & 4 &     0.5 &     0.000474 &  1.3525E-02 &   2.002313 & + &  e & X2Sigma+ &     1 &   0 &  0.5 &  0.5 & Ma &  1260.131222\\
3 & 2481.999341 & 4 &     0.5 &     0.000468 &  7.2418E-03 &   2.002313 & + &  e & X2Sigma+ &     2 &   0 &  0.5 &  0.5 & Ma &  2481.994238\\
4 & 3665.414217 & 4 &     0.5 &     0.000549 &  5.1897E-03 &   2.002313 & + &  e & X2Sigma+ &     3 &   0 &  0.5 &  0.5 & Ma &  3665.412944\\
5 & 4809.943468 & 4 &     0.5 &     2.507500 &  4.2007E-03 &   2.002313 & + &  e & X2Sigma+ &     4 &   0 &  0.5 &  0.5 & Ca &  4809.943468 \\ 
\bottomrule\bottomrule
\end{tabular}}
\mbox{}\\
{\flushleft
$i$: State counting number.     \\
$\tilde{E}$: State energy (in cm$^{-1}$). \\
$g_i$:  Total statistical weight, equal to ${g_{\rm ns}(2J + 1)}$.     \\
$J$: Total angular momentum.\\
unc: Uncertainty (in cm$^{-1}$).\\
$\tau$: Lifetime (in seconds).\\
$g$: Land\'{e} $g$-factor. \\
$+/-$:   Total parity. \\
$e/f$:   Rotationless parity. \\
State: Electronic state.\\
$v$:   State vibrational quantum number. \\
$\Lambda$:  Projection of the electronic angular momentum. \\
$\Sigma$:   Projection of the electronic spin. \\
$\Omega$:   Projection of the total angular momentum, $\Omega=\Lambda+\Sigma$. \\
Label: ``Ma'' for MARVEL, ``Ca'' for calculated. \\
Calc: Original \Duo\ calculated state energy (in cm$^{-1}$).\\
}
\end{table*}

\subsection{Temperature-dependent partition functions}
\label{sec:pfn}

Calculations of the temperature-dependent partition function $Q(T)$, defined as
\begin{equation}
\label{eq:pfn}
Q(T)=\sum_{i} g_i \exp\left(\frac{-E_i}{kT}\right) ,
\end{equation}
were performed on a $1$~K grid in the 1\,--\,5000~K range for CaH and MgH and its isotopologues (provided as supplementary material). Here, $g_i=g_{\rm ns}(2J_i+1)$ is the degeneracy of a state $i$ with energy $E_i$ and rotational quantum number $J_i$. For the nuclear spin statistical weights we have used values of $g_{\rm ns}=2$ for $^{40}$CaH, $^{24}$MgH and $^{26}$MgH, and $g_{\rm ns}=12$ for $^{25}$MgH. Note that we have chosen to go up to 5000~K as $B$--$X$ lines of CaH have been observed in umbral spectra of sunspots with an average rotational temperature of $4164\pm 164$~K~\citep{d20BeDeSu.CaH}.

\begin{table}
\centering
\caption{\label{tab:pf}Computed values of the partition function $Q(T)$ for different temperatures $T$ (in K) compared against \citet{jt529} for $^{40}$CaH and against \citet{15SzCsxx.MgH} (bound plus resonance state values; see text) for $^{24}$MgH.}
\begin{tabular}{ccccccc}
\toprule\toprule
\multirow{2}{*}{} & & \multicolumn{1}{c}{$^{40}$CaH} & & & \multicolumn{1}{c}{$^{24}$MgH} &\\
\cmidrule(l){2-4} \cmidrule(l){5-7}
$T$ & $Q$(This work) & $Q$\citep{jt529} & \% difference &
$Q$(This work) & $Q$\citep{15SzCsxx.MgH} & \% difference\\
\midrule
1000 & 804.5 & 804.6 & 0.0 & 569.7 & 569.7 & 0.0\\
2000 & 2351.8 & 2352.8 & 0.0 & 1631.1 & 1622.9 & 0.5\\
3000 & 4844.2 & 4885.7 & 0.9 & 3463.1 & 3337.7 & 3.6\\
\bottomrule\bottomrule
\end{tabular}
\end{table}

In Table~\ref{tab:pf}, computed partition function values at different temperatures are shown for $^{40}$CaH and $^{24}$MgH. For CaH, we have compared against the $Q(T)$ values from the previous ExoMol study~\citep{jt529}, which only considered energy levels in the \X\ ground state in the summation of Eq.~\eqref{eq:pfn}. There is very little difference in values, demonstrating that the inclusion of excited electronic state energy levels has an almost negligible contribution to $Q(T)$, even at higher temperatures. Our partition function values are actually slightly smaller than \citet{jt529} as we use a lower dissociation asymptote for the \X\ state PEC and only include vibrational levels up to $v_{\mathrm{max}}=15$ (as oppose to $v_{\mathrm{max}}=19$).

For MgH we have compared against the study of \citet{15SzCsxx.MgH}, which rigorously treated the contribution from bound, resonance (quasibound) and unbound states in accurate calculations of $Q(T)$ up to $T=3000$~K. Since we have considered a large number of quasibound states in our calculations, we have taken the $Q(T)$ values that included bound and resonance states from \citet{15SzCsxx.MgH}. Our computed partition function values are slightly larger at higher temperatures but this difference is relatively small. Interestingly, if only including bound levels of the \X\ state (below $\approx 10\,400$~cm$^{-1}$) in the summation of Eq.~\eqref{eq:pfn} our computed partition function values are near-identical to the bound state values of \citet{15SzCsxx.MgH}.

\subsection{Simulated spectra of CaH and MgH}
\label{sec:spectra_cah_mgh}

\subsubsection{CaH}

\begin{figure}
\centering
\includegraphics[width=0.7\textwidth]{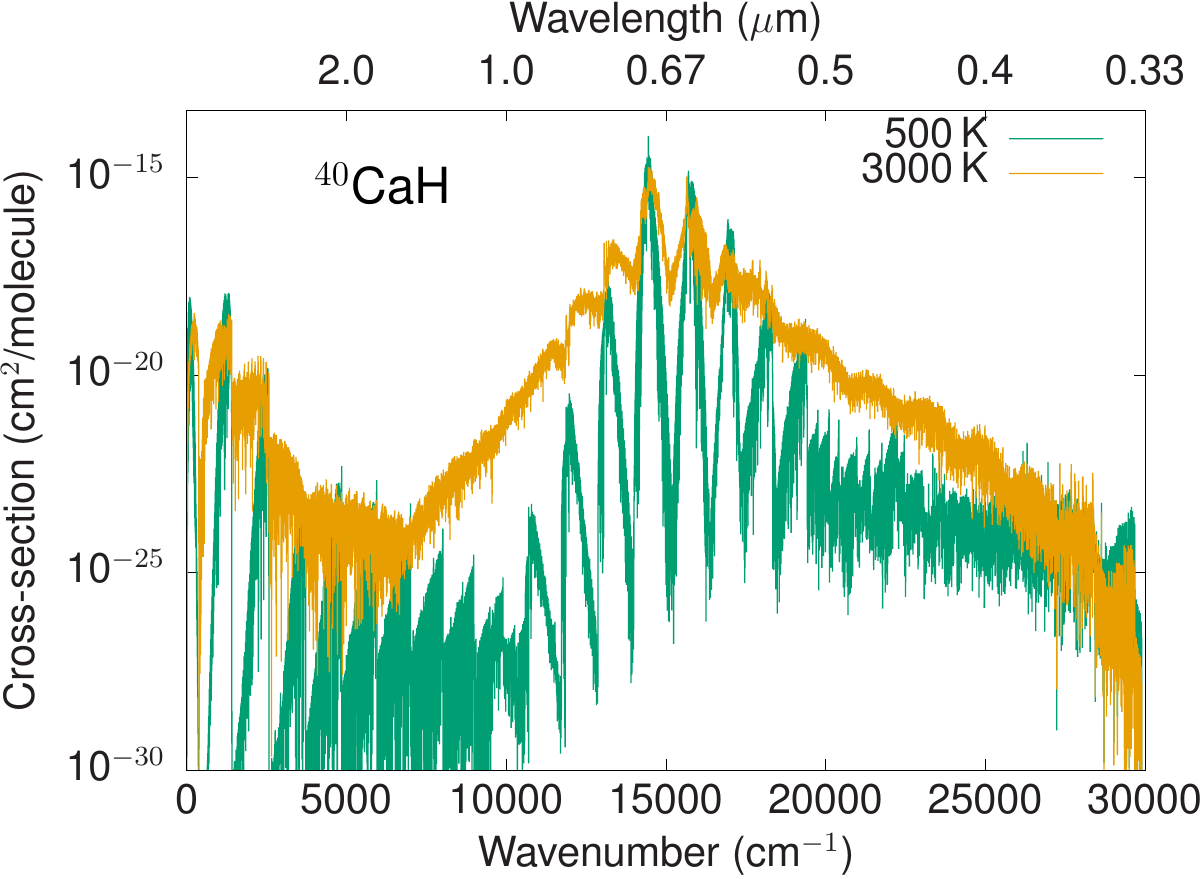}
\caption{\label{fig:500K-3000K_cah}Spectrum of $^{40}$CaH at $T=500$~K and $T=3000$~K. Absorption cross-sections were computed at a resolution of 1~cm$^{-1}$ and modelled with a Gaussian line profile with a half width at half maximum (HWHM) of 1~cm$^{-1}$.}
\end{figure}

\begin{figure}
\centering
\includegraphics[width=0.7\textwidth]{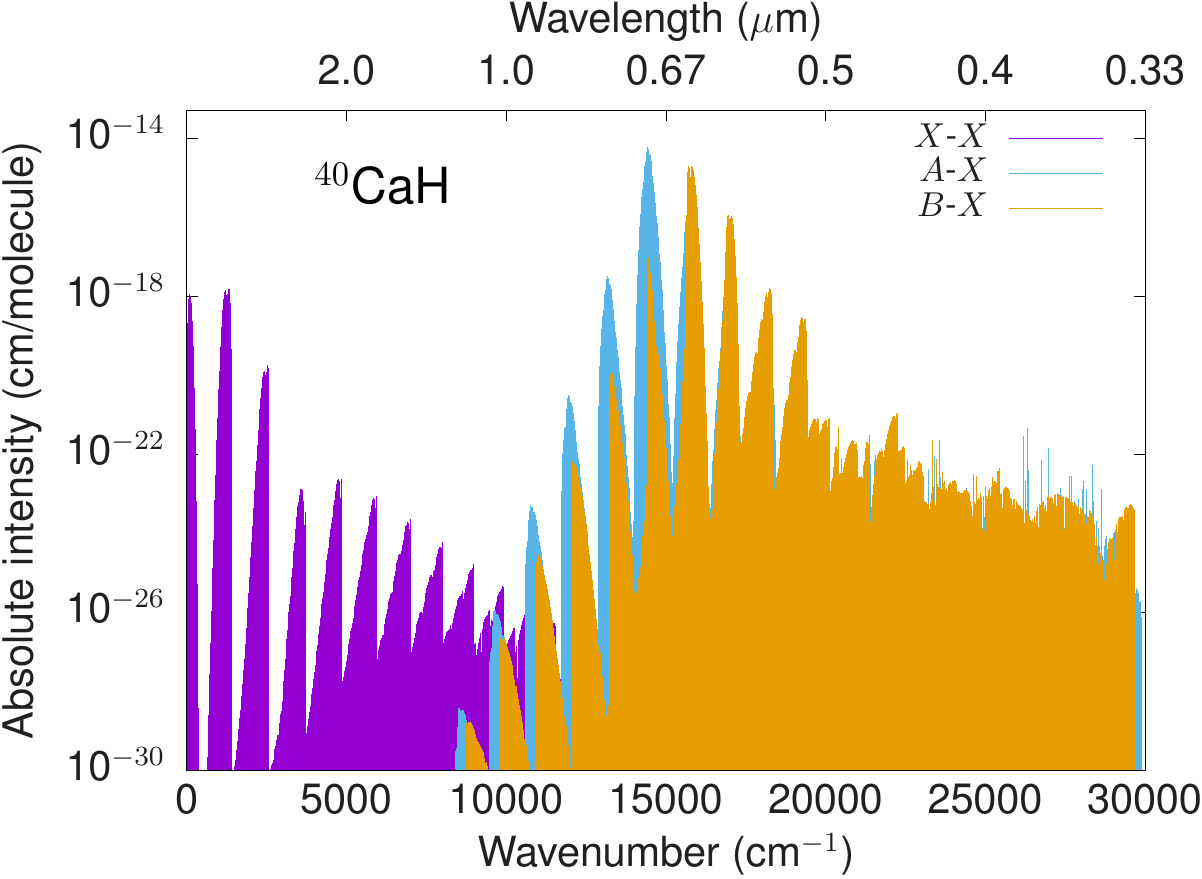}
\caption{\label{fig:XAB_cah}Absolute absorption line intensities of the $X$--$X$, $A$--$X$ and $B$--$X$ bands of $^{40}$CaH simulated at $T=500$~K.}
\end{figure}

All spectral simulations were performed with the \textsc{ExoCross} program~\citep{jt708}. In Fig.~\ref{fig:500K-3000K_cah}, an overview of the rovibronic spectrum of CaH is displayed where we have simulated integrated absorption cross-sections at a resolution of 1~cm$^{-1}$ modelled with a Gaussian line profile with a half width at half maximum (HWHM) of 1~cm$^{-1}$. Spectra have been generated at $T=500$~K and $T=3000$~K to illustrate the spectral flattening that occurs at higher temperatures as weaker features gain more intensity due to the increased population of higher-energy rovibronic states. The strongest band around $\approx 14\,450$~cm$^{-1}$, (largely due to $A$--$X$ transitions) still dominates at higher temperatures along with the second strongest band around $\approx 15\,750$~cm$^{-1}$ (largely due to $B$--$X$ transitions). The different contributions to the spectrum from $X$--$X$, $A$--$X$ and $B$--$X$ transitions is illustrated in Fig.~\ref{fig:XAB_cah}.

\begin{figure}
\centering
\includegraphics[width=0.5\textwidth]{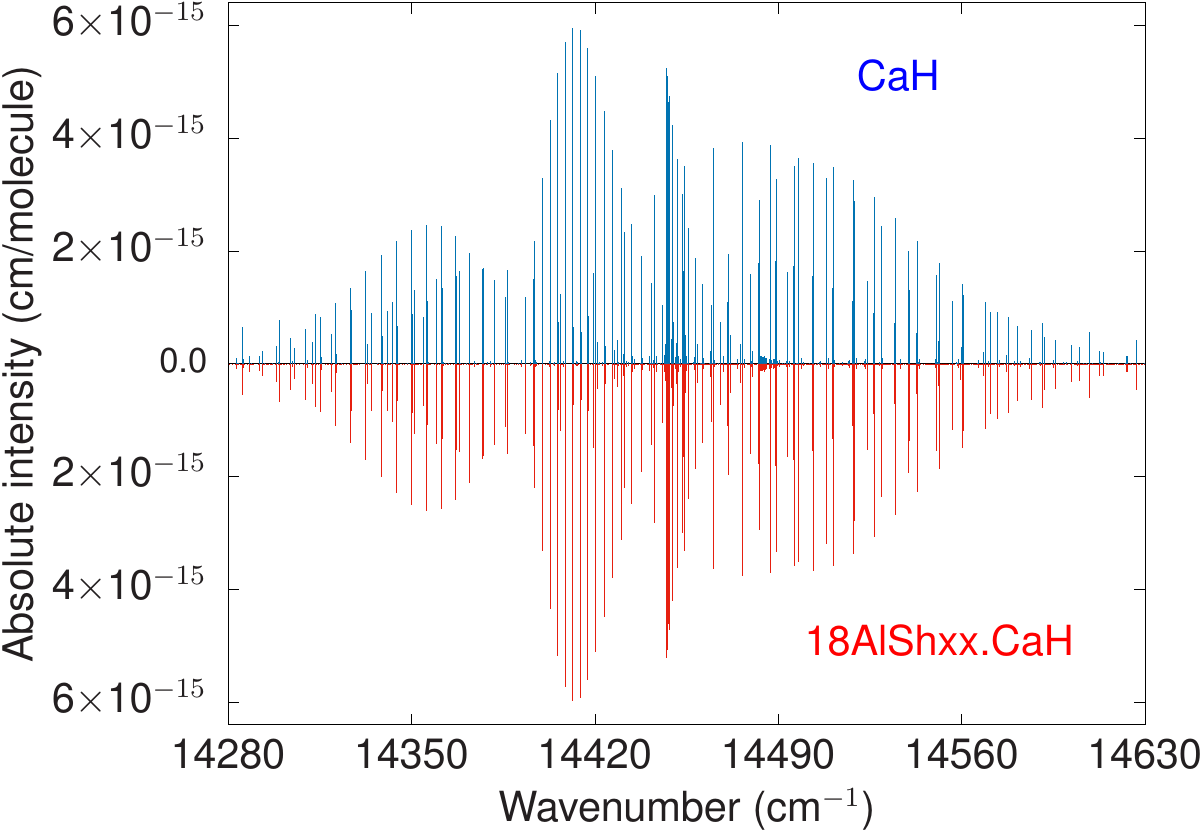}
\includegraphics[width=0.5\textwidth]{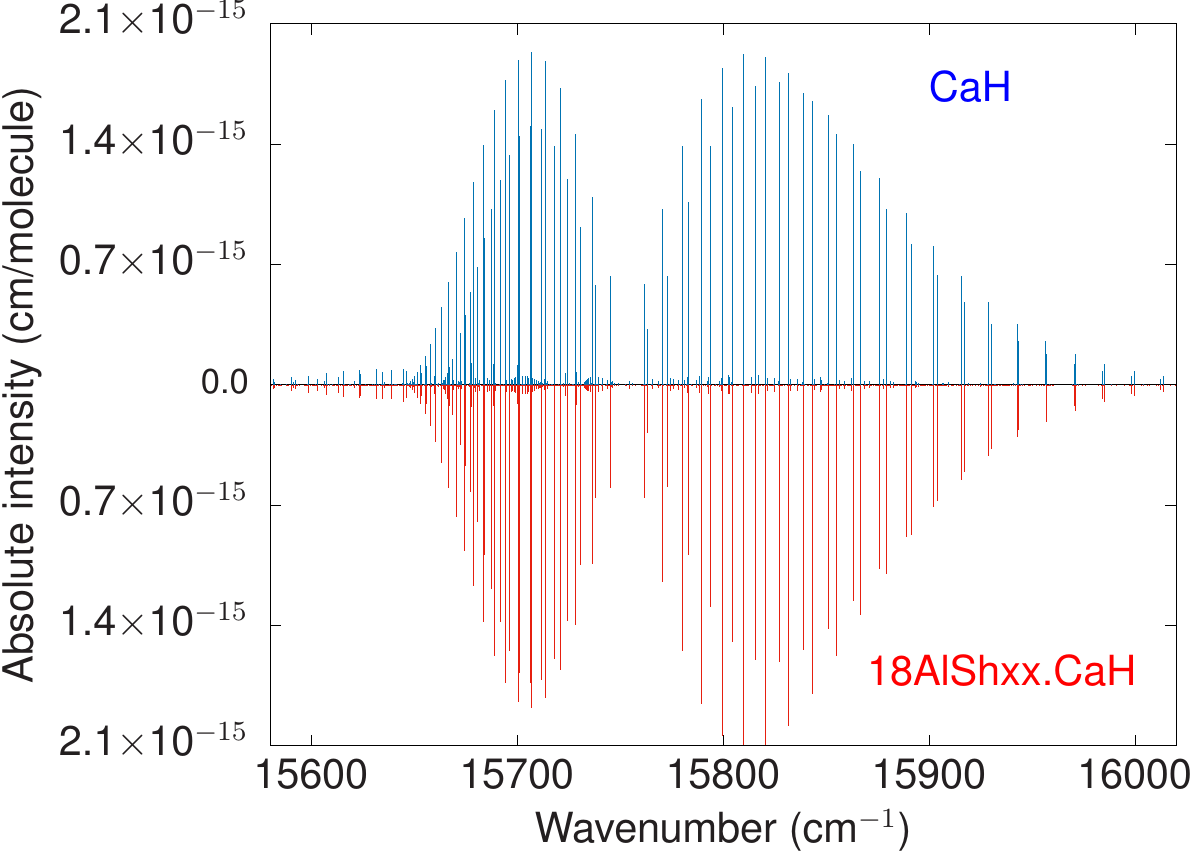}
\includegraphics[width=0.5\textwidth]{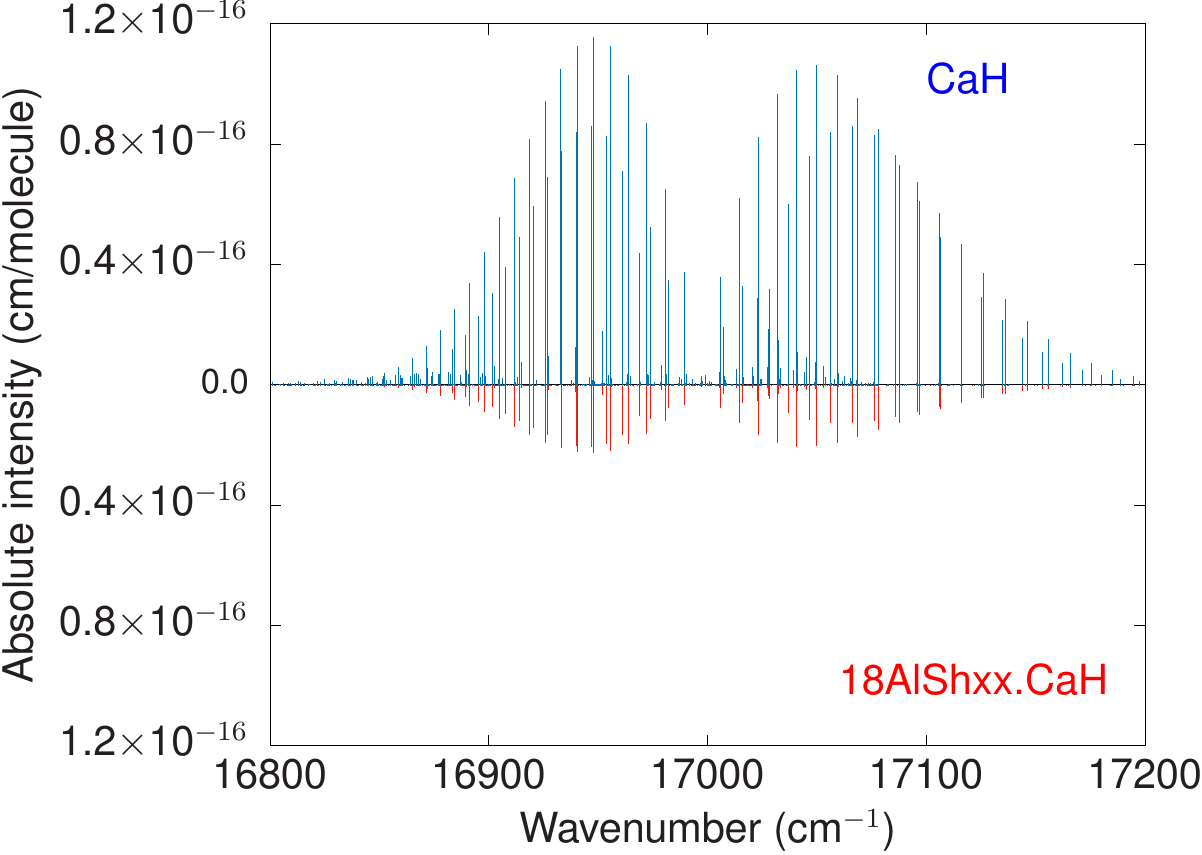}
\caption{\label{fig:500K_shayesteh_cah}Absolute absorption line intensities of the $A$--$X$ and $B$--$X$ bands of CaH simulated at $T=500$~K compared against the rovibronic line list of 18AlShxx.CaH~\citep{18AlShxx.CaH}.}
\end{figure}

Absolute absorption line intensities have been calculated at $T=500$~K and are compared against the rovibronic line list of \citet{18AlShxx.CaH} in Fig.~\ref{fig:500K_shayesteh_cah}. There is very good agreement with the line list of \citet{18AlShxx.CaH} and this is to be expected since we utilise the same \textit{ab initio} TDMCs and would therefore expect similar line intensities. Interestingly, however, for a relatively weaker band located around 17\,000~cm$^{-1}$ our computed line intensities are much stronger than those of \citet{18AlShxx.CaH}, as seen in the last panel of Fig.~\ref{fig:500K_shayesteh_cah}. Given that we use the same transition dipoles, this difference can be attributed to the nuclear wavefunctions used in the calculation of the Einstein $A$ coefficients. Since our spectroscopic model of CaH rigorously treats couplings between electronic states, we expect our intensities to be more reliable as a result of the improved nuclear wavefunctions.

\subsubsection{MgH}

\begin{figure}
\centering
\includegraphics[width=0.7\textwidth]{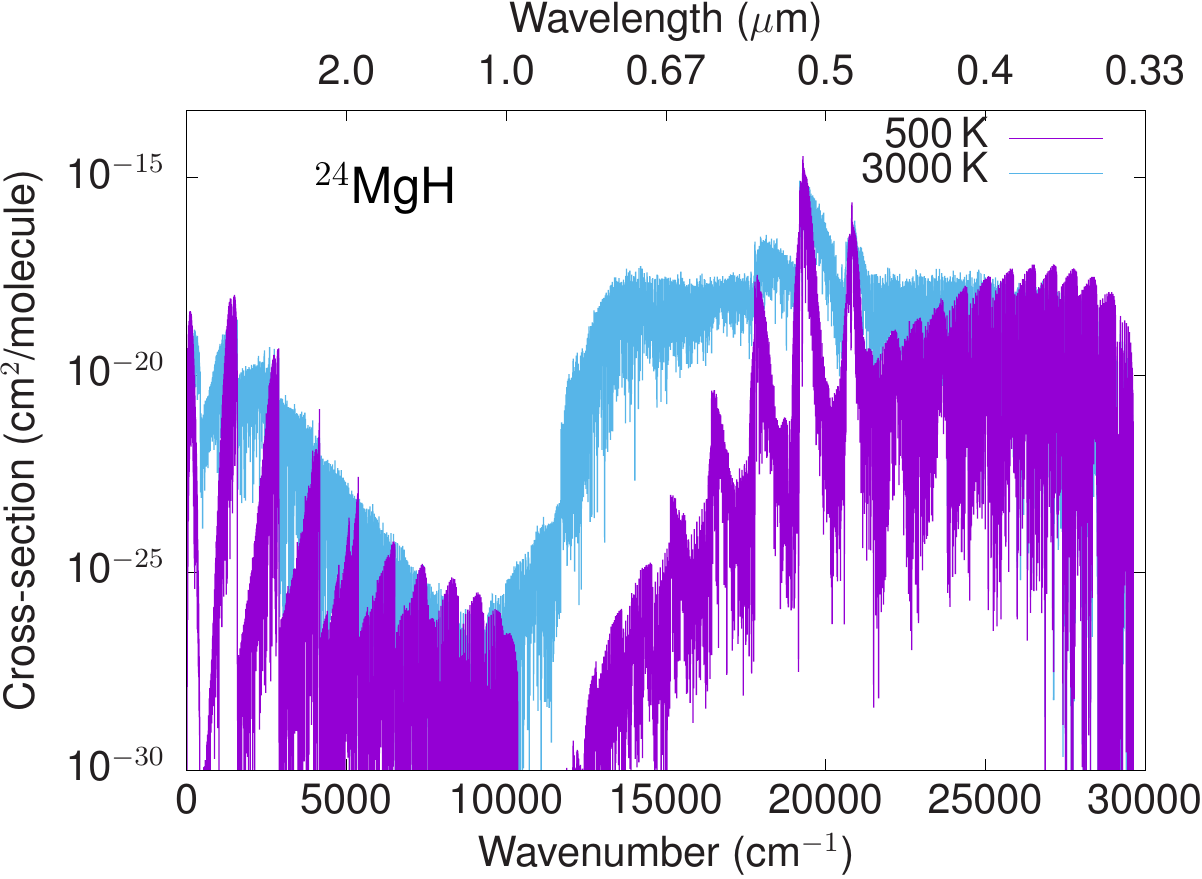}
\caption{\label{fig:500K-3000K_mgh}Spectrum of $^{24}$MgH at $T=500$~K and $T=3000$~K. Absorption cross-sections were computed at a resolution of 1~cm$^{-1}$ and modelled with a Gaussian line profile with a half width at half maximum (HWHM) of 1~cm$^{-1}$.}
\end{figure}

\begin{figure}
\centering
\includegraphics[width=0.7\textwidth]{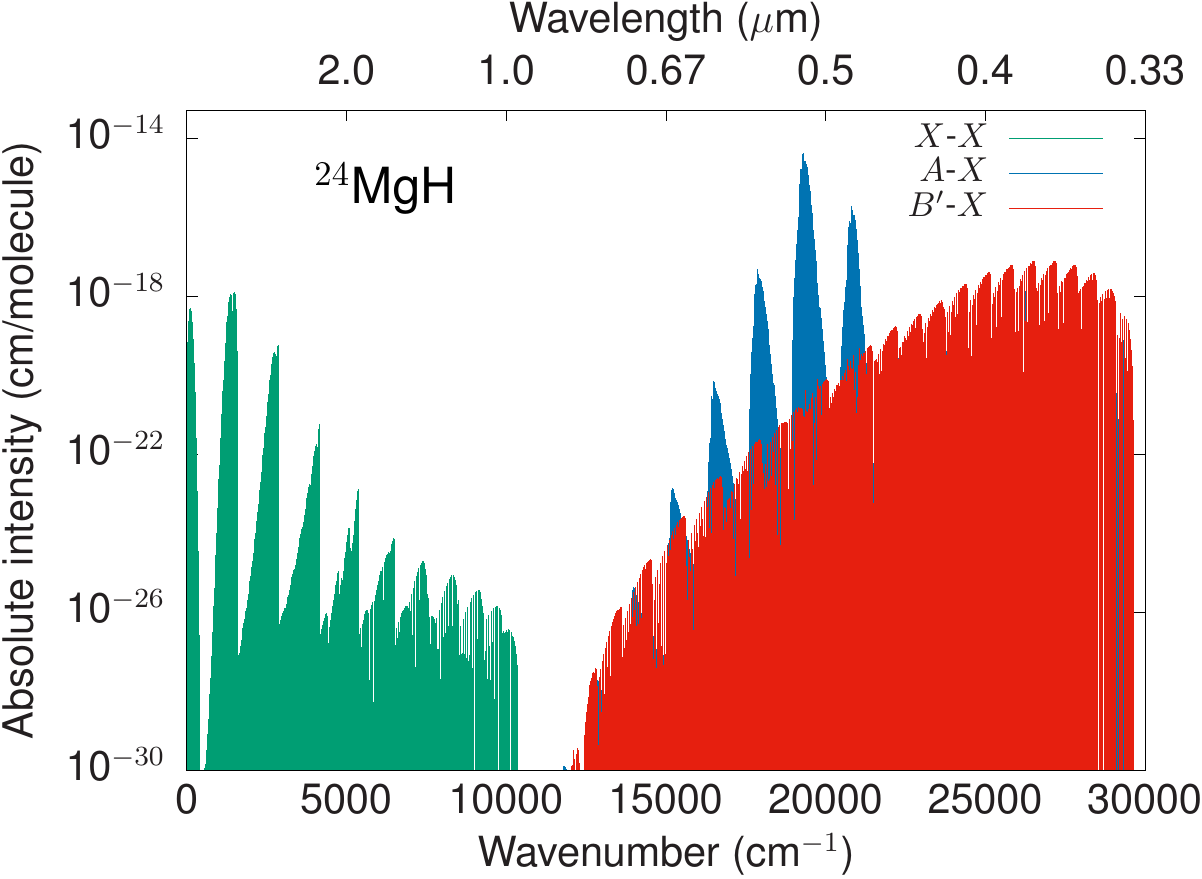}
\caption{\label{fig:XAB_mgh}Absolute absorption line intensities of the $X$--$X$, $A$--$X$ and $B^{\prime}$--$X$ bands of $^{24}$MgH simulated at $T=500$~K.}
\end{figure}

\begin{figure}
\centering
\includegraphics[width=0.5\textwidth]{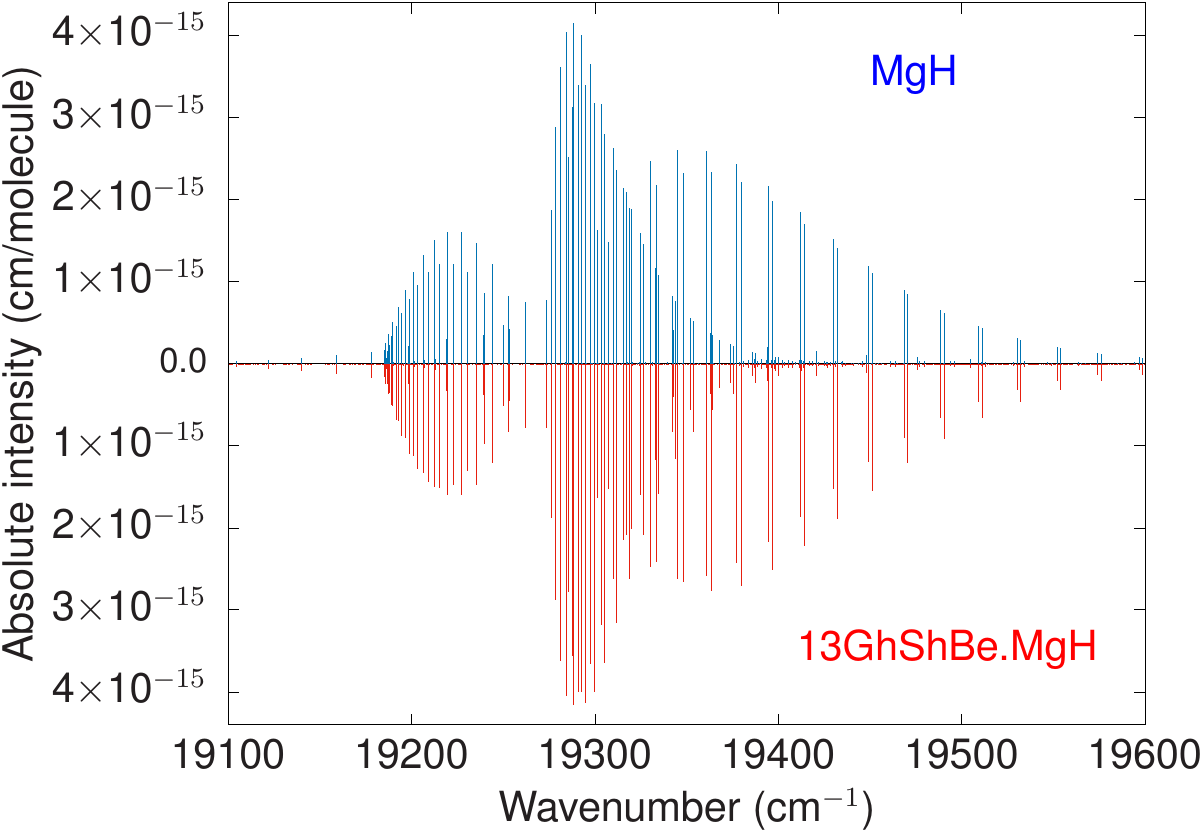}
\includegraphics[width=0.5\textwidth]{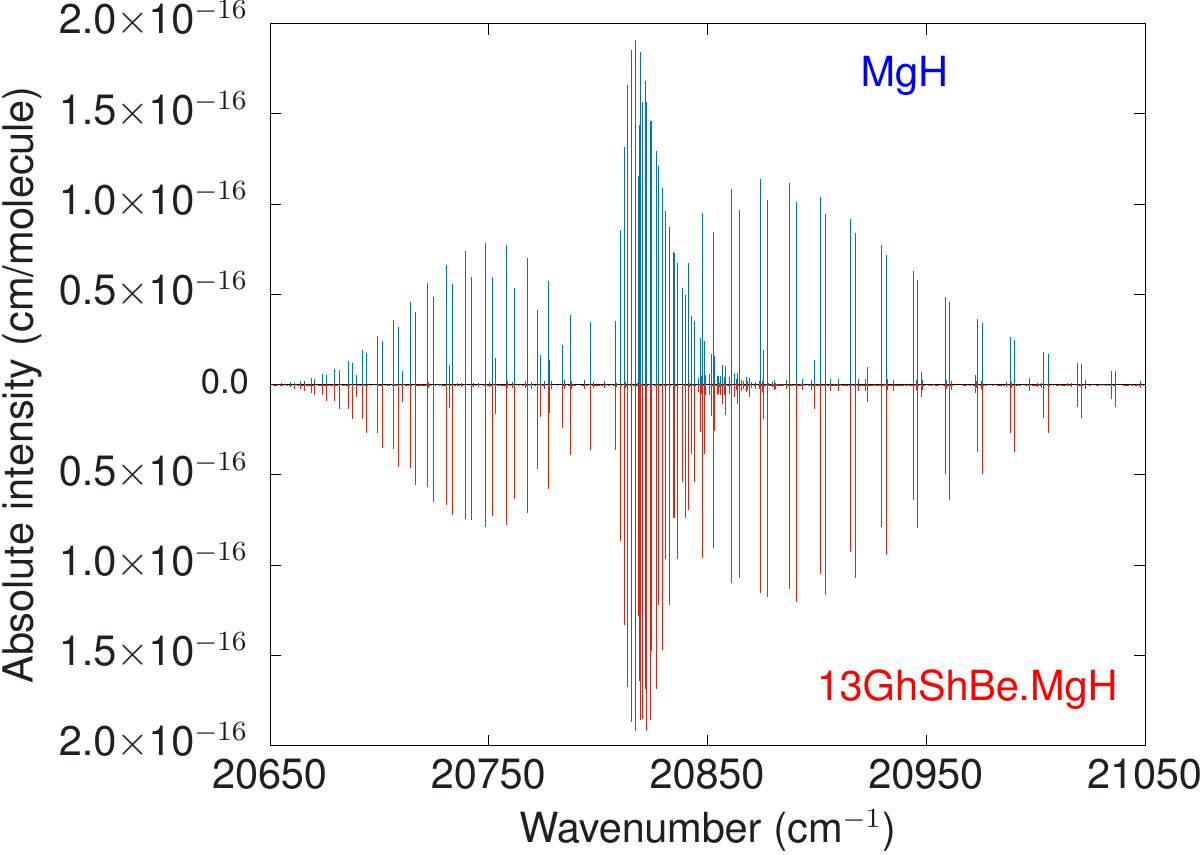}
\caption{\label{fig:500K_bernath_mgh}Absolute absorption line intensities of the $A$--$X$ and $B^{\prime}$--$X$ bands of $^{24}$MgH simulated at $T=500$~K compared against the rovibronic line list of 13GhShBe.MgH~\citep{13GhShBe.MgH}.}
\end{figure}

The temperature-dependence of the $^{24}$MgH spectrum is illustrated in Fig.~\ref{fig:500K-3000K_mgh} where we have plotted absorption cross-sections at $T=500$~K and $T=3000$~K (same resolution and line profile as CaH cross-sections), while the different contributions to the spectrum from $X$--$X$, $A$--$X$ and $B^{\prime}$--$X$ transitions is shown in Fig.~\ref{fig:XAB_mgh}. The $B^{\prime}$--$X$ contribution is completely different in shape compared to the $A$--$X$ bands, explained by the fact that the minimum of the \Bp\ PEC lies at a larger equilibrium bond length compared to the \X\ and \A\ PECs (see Fig.~\ref{fig:pec_mgh}). Absolute line intensities of the two strongest bands at $T=500$~K are shown in Fig.~\ref{fig:500K_bernath_mgh} and compared against the rovibronic line list of \citet{13GhShBe.MgH}. There is excellent agreement for both line positions and intensities, which again is to be expected since we use the same \textit{ab initio} transition dipoles and our PECs and couplings were refined to a lot of the experimental transition data used by \citet{13GhShBe.MgH} to determine their PECs. Spectra of the isotopologues $^{25}$MgH and $^{26}$MgH are very similar to the main $^{24}$MgH isotopologue so we do not show any plots here. These line lists will be useful, for example, in establishing magnesium isotope ratios in stellar environments~\citep{03DaYDaL.MgH}.

\section{Conclusion}
\label{sec:conc}

New line lists for calcium monohydride ($^{40}$CaH) and magnesium monohydride ($^{24}$MgH) and its minor isotopologues ($^{25}$MgH, $^{26}$MgH) have been presented. The line lists cover the 0--30\,000~cm$^{-1}$ region (wavelengths $\lambda > 0.33$~$\mu$m) and are applicable to  temperatures up to 5000~K. Compared to previous CaH and MgH rovibronic line lists, notably the most recent~\citep{18AlShxx.CaH,13GhShBe.MgH}, the new ExoMol line lists rigorously treat the effects of coupling in/between electronic states and have improved coverage as they consider energy levels with rovibrational excitation up to the dissociation limit of each electronic state, importantly in the excited \A\ and \BBp\ states. A large number of quasibound levels in the \X\ ground state of MgH were also included in our calculations as they influence the spectrum at higher temperatures~\citep{07ShHeLe.MgH}. Hence we name
these line lists \texttt{XAB}.

At microwave and IR wavelengths the new CaH and MgH line lists are recommended instead of the previous ExoMol 2012 line lists~\citep{jt529}. The replacement of calculated energy levels with empirical-quality MARVEL values will vastly improve the accuracy in certain regions making them suitable for high-resolution observations, especially in the IR region as a large number of \X\ state energy levels were substituted in both molecules. The calculation of Land\'e $g$-factors should also be of use to applications involving these molecules in the presence of weak magnetic fields. There have been laboratory Zeeman spectroscopic studies of CaH~\citep{06ChGeSt.CaH} and MgH~\citep{14ZhStxx.MgH} and there is motivation for using these molecules to probe stellar magnetic fields~\citep{15AfBexx}, notably using the $A$--$X$ band.

\section*{Acknowledgments}

This work was supported by the STFC Projects No. ST/M001334/1 and ST/R000476/1. The authors acknowledge the use of the UCL Legion High Performance Computing Facility (Legion@UCL) and associated support services in the completion of this work, along with the Cambridge Service for Data Driven Discovery (CSD3), part of which is operated by the University of Cambridge Research Computing on behalf of the STFC DiRAC HPC Facility (www.dirac.ac.uk). The DiRAC component of CSD3 was funded by BEIS capital funding via STFC capital grants ST/P002307/1 and ST/R002452/1 and STFC operations grant ST/R00689X/1. DiRAC is part of the National e-Infrastructure.
This work was also supported by the European
Research Council (ERC) under the European Union’s Horizon 2020 research and innovation
programme through Advance Grant number 883830.

\section*{Data Availability}

The \textsc{Duo} model input files plus the  MARVEL transitions and energy files are given
 as supplementary material to this article.
The  states, transition and partition function files for the \texttt{XAB} line lists can be downloaded from
\href{www.exomol.com}{www.exomol.com} and the CDS data centre
\href{http://cdsarc.u-strasbg.fr}{cdsarc.u-strasbg.fr}. The open access  programs \textsc{ExoCross} and \textsc{Duo} are  available from \href{https://github.com/exomol}{github.com/exomol}.

\section*{Supporting Information}
Supplementary data are available at MNRAS online. This includes 
a detailed description of the spectroscopic models of CaH and MgH, input files for the program \Duo\ containing the spectroscopic models of CaH and MgH, and MARVEL transitions and energy files. The following references were cited in the supplementary material: \citet{MLRpaper,jt529,EMO,99MeHuxx.methods,99SkPeBo.methods,79BrMexx.methods,13HeShTa.MgH,11ShBexx.MgH,12MoShxx.MgH}

\bibliographystyle{mnras}

\begin{thebibliography}{}
\makeatletter
\relax
\def\mn@urlcharsother{\let\do\@makeother \do\$\do\&\do\#\do\^\do\_\do\%\do\~}
\def\mn@doi{\begingroup\mn@urlcharsother \@ifnextchar [ {\mn@doi@}
  {\mn@doi@[]}}
\def\mn@doi@[#1]#2{\def\@tempa{#1}\ifx\@tempa\@empty \href
  {http://dx.doi.org/#2} {doi:#2}\else \href {http://dx.doi.org/#2} {#1}\fi
  \endgroup}
\def\mn@eprint#1#2{\mn@eprint@#1:#2::\@nil}
\def\mn@eprint@arXiv#1{\href {http://arxiv.org/abs/#1} {{\tt arXiv:#1}}}
\def\mn@eprint@dblp#1{\href {http://dblp.uni-trier.de/rec/bibtex/#1.xml}
  {dblp:#1}}
\def\mn@eprint@#1:#2:#3:#4\@nil{\def\@tempa {#1}\def\@tempb {#2}\def\@tempc
  {#3}\ifx \@tempc \@empty \let \@tempc \@tempb \let \@tempb \@tempa \fi \ifx
  \@tempb \@empty \def\@tempb {arXiv}\fi \@ifundefined
  {mn@eprint@\@tempb}{\@tempb:\@tempc}{\expandafter \expandafter \csname
  mn@eprint@\@tempb\endcsname \expandafter{\@tempc}}}

\bibitem[\protect\citeauthoryear{Afram \& Berdyugina}{Afram \&
  Berdyugina}{2015}]{15AfBexx}
Afram N.,  Berdyugina S.~V.,  {2015}, \mn@doi [A\&A]
  {{10.1051/0004-6361/201425314}}, {576}, A34

\bibitem[\protect\citeauthoryear{Alavi \& Shayesteh}{Alavi \&
  Shayesteh}{2018}]{18AlShxx.CaH}
Alavi S.~F.,  Shayesteh A.,  2018, \mn@doi [MNRAS] {10.1093/mnras/stx2681},
  474, 2

\bibitem[\protect\citeauthoryear{Balfour \& Lindgren}{Balfour \&
  Lindgren}{1978}]{78BaLixx.MgH}
Balfour W.~J.,  Lindgren B.,  1978, Can. J. Phys., 56, 767

\bibitem[\protect\citeauthoryear{Behere, Deshmukh, Patil  \& Behere}{Behere
  et~al.}{2020}]{d20BeDeSu.CaH}
Behere S.,  Deshmukh B.,  Patil S.,   Behere S.~H.,  2020, \mn@doi [AIP Conf.
  Proc.] {10.1063/5.0005460}, 2224, 020001

\bibitem[\protect\citeauthoryear{Bell, Edvardsson  \& Gustafsson}{Bell
  et~al.}{1985}]{85BeEdGu.MgH}
Bell R.~A.,  Edvardsson B.,   Gustafsson B.,  1985, MNRAS, 212, 497

\bibitem[\protect\citeauthoryear{Berg, Klynning  \& Martin}{Berg
  et~al.}{1976}]{76BeKlMa.CaH}
Berg L.~E.,  Klynning L.,   Martin H.,  1976, \mn@doi [Opt. Commun.]
  {10.1016/0030-4018(76)90270-4}, 17, 320

\bibitem[\protect\citeauthoryear{Bernath}{Bernath}{2020}]{MOLLIST}
Bernath P.~F.,  2020, \mn@doi [J. Quant. Spectrosc. Radiat. Transf.]
  {10.1016/j.jqsrt.2019.106687}, 240, 106687

\bibitem[\protect\citeauthoryear{Birkby}{Birkby}{2018}]{18Birkby}
Birkby J.~L.,  2018, Handbook of Exoplanets, pp 1485--1508

\bibitem[\protect\citeauthoryear{Bonnell \& Bell}{Bonnell \&
  Bell}{1993}]{93BoBexx.MgH}
Bonnell J.~T.,  Bell R.~A.,  1993, MNRAS, 264, 334

\bibitem[\protect\citeauthoryear{Bowesman, Shuai, Yurchenko  \&
  Tennyson}{Bowesman et~al.}{2021}]{jt835}
Bowesman C.~A.,  Shuai M.,  Yurchenko S.~N.,   Tennyson J.,  2021, \mn@doi
  [MNRAS] {10.1093/mnras/stab2525}, 508, 3181

\bibitem[\protect\citeauthoryear{Brown \& Merer}{Brown \&
  Merer}{1979}]{79BrMexx.methods}
Brown J.~M.,  Merer A.~J.,  1979, \mn@doi [J. Mol. Spectrosc.]
  {10.1016/0022-2852(79)90172-3}, 74, 488

\bibitem[\protect\citeauthoryear{Chen, Gengler, Steimle  \& Brown}{Chen
  et~al.}{2006}]{06ChGeSt.CaH}
Chen J.~H.,  Gengler J.,  Steimle T.~C.,   Brown J.~M.,  2006, \mn@doi [Phys.
  Rev. A] {10.1103/PhysRevA.73.012502}, 73, 012502

\bibitem[\protect\citeauthoryear{Chubb et~al.,}{Chubb et~al.}{2021}]{jt801}
Chubb K.~L.,  et~al., 2021, \mn@doi [A\&A] {10.1051/0004-6361/202038350}, 646,
  A21

\bibitem[\protect\citeauthoryear{Conroy, Villaume, van Dokkum  \& Lind}{Conroy
  et~al.}{2018}]{Conroy:2018}
Conroy C.,  Villaume A.,  van Dokkum P.~G.,   Lind K.,  2018, \mn@doi [ApJ]
  {10.3847/1538-4357/aaab49}, 854, 139

\bibitem[\protect\citeauthoryear{Cs{\'a}sz{\'a}r, Czak{\'o}, Furtenbacher  \&
  M{\'a}tyus}{Cs{\'a}sz{\'a}r et~al.}{2007}]{07CsCzFu.method}
Cs{\'a}sz{\'a}r A.~G.,  Czak{\'o} G.,  Furtenbacher T.,   M{\'a}tyus E.,  2007,
  Annu. Rep. Comput. Chem., 3, 155

\bibitem[\protect\citeauthoryear{Di~Rosa}{Di~Rosa}{2004}]{04DiRosa.CaH}
Di~Rosa M.~D.,  2004, \mn@doi [Eur. Phys. J. D] {10.1140/epjd/e2004-00167-2},
  31, 395

\bibitem[\protect\citeauthoryear{Dunning}{Dunning}{1989}]{89Dunning.ai}
Dunning T.~H.,  1989, \mn@doi [J. Chem. Phys.] {10.1063/1.456153}, 90, 1007

\bibitem[\protect\citeauthoryear{Fowler}{Fowler}{1907}]{07Fowler.MgH}
Fowler A.,  1907, \mn@doi [MNRAS] {10.1093/mnras/67.8.530}, 67, 530

\bibitem[\protect\citeauthoryear{Fowler}{Fowler}{1909}]{09Fowler.MgH}
Fowler A.,  1909, \mn@doi [Phil. Trans. Royal Soc. London A]
  {10.1098/rsta.1909.0017}, 209, 447

\bibitem[\protect\citeauthoryear{Furtenbacher \& {Cs\'asz\'ar}}{Furtenbacher \&
  {Cs\'asz\'ar}}{2012}]{12FuCsxx.methods}
Furtenbacher T.,  {Cs\'asz\'ar} A.~G.,  2012, \mn@doi [J. Mol. Struct.]
  {10.1016/j.molstruc.2011.10.057}, 1009, 123

\bibitem[\protect\citeauthoryear{Furtenbacher, {Cs\'asz\'ar}  \&
  Tennyson}{Furtenbacher et~al.}{2007}]{jt412}
Furtenbacher T.,  {Cs\'asz\'ar} A.~G.,   Tennyson J.,  2007, \mn@doi [J. Mol.
  Spectrosc.] {10.1016/j.jms.2007.07.005}, 245, 115

\bibitem[\protect\citeauthoryear{Gandhi \& Madhusudhan}{Gandhi \&
  Madhusudhan}{2019}]{Gandhi:2019}
Gandhi S.,  Madhusudhan N.,  2019, \mn@doi [MNRAS] {10.1093/mnras/stz751}, 485,
  5817

\bibitem[\protect\citeauthoryear{Gandhi et~al.,}{Gandhi et~al.}{2020}]{jt782}
Gandhi S.,  et~al., 2020, \mn@doi [MNRAS] {10.1093/mnras/staa981}, 495,
  224–237

\bibitem[\protect\citeauthoryear{Gao}{Gao}{2020}]{20Gaoxxx.MgH}
Gao Y.,  {2020}, \mn@doi [Phys. Rev. A] {{10.1103/PhysRevA.102.042821}}, {102},
  042821

\bibitem[\protect\citeauthoryear{Gao \& Gao}{Gao \& Gao}{2014}]{14GaGaxx.MgH}
Gao Y.,  Gao T.,  2014, \mn@doi [Phys. Rev. A] {10.1103/PhysRevA.90.052506},
  90, 052506

\bibitem[\protect\citeauthoryear{Gay \& Lambert}{Gay \&
  Lambert}{2000}]{00GaLaxx.MgH}
Gay P.~L.,  Lambert D.~L.,  2000, \mn@doi [ApJ] {10.1086/308653}, 533, 260

\bibitem[\protect\citeauthoryear{Gharib-Nezhad, Iyer, Line, Freedman, Marley
  \& Batalha}{Gharib-Nezhad et~al.}{2021}]{21GhIyLi}
Gharib-Nezhad E.,  Iyer A.~R.,  Line M.~R.,  Freedman R.~S.,  Marley M.~S.,
  Batalha N.~E.,  {2021}, \mn@doi [ApJS] {{10.3847/1538-4365/abf504}}, {254},
  34

\bibitem[\protect\citeauthoryear{GharibNezhad, Shayesteh  \&
  Bernath}{GharibNezhad et~al.}{2013}]{13GhShBe.MgH}
GharibNezhad E.,  Shayesteh A.,   Bernath P.~F.,  {2013}, \mn@doi [MNRAS]
  {10.1093/mnras/stt510}, {432}, 2043

\bibitem[\protect\citeauthoryear{Grimm et~al.,}{Grimm et~al.}{2021}]{jt819}
Grimm S.~L.,  et~al., 2021, \mn@doi [ApJS] {10.3847/1538-4365/abd773}, 253, 30

\bibitem[\protect\citeauthoryear{Hawkins, Masseron, Jofre, Gilmore, Elsworth
  \& Hekker}{Hawkins et~al.}{2016}]{Hawkins:2016}
Hawkins K.,  Masseron T.,  Jofre P.,  Gilmore G.,  Elsworth Y.,   Hekker S.,
  2016, \mn@doi [A\&A] {10.1051/0004-6361/201628812}, 594, A43

\bibitem[\protect\citeauthoryear{Henderson, Shayesteh, Tao, Haugen, Bernath  \&
  Le~Roy}{Henderson et~al.}{2013}]{13HeShTa.MgH}
Henderson R. D.~E.,  Shayesteh A.,  Tao J.,  Haugen C.~C.,  Bernath P.~F.,
  Le~Roy R.~J.,  {2013}, \mn@doi [J. Phys. Chem. A] {10.1021/jp406680r}, {117},
  13373

\bibitem[\protect\citeauthoryear{Hinkle, Wallace, Ram, Bernath, Sneden  \&
  Lucatello}{Hinkle et~al.}{2013}]{13HiWaRa.MgH}
Hinkle K.~H.,  Wallace L.,  Ram R.~S.,  Bernath P.~F.,  Sneden C.,   Lucatello
  S.,  2013, \mn@doi [ApJS] {10.1088/0067-0049/207/2/26}, 207, 26

\bibitem[\protect\citeauthoryear{Jun-Hao, Tao  \& Jian-Ping}{Jun-Hao
  et~al.}{2021}]{21YiYaYi.CaH}
Jun-Hao Y.,  Tao Y.,   Jian-Ping Y.,  {2021}, \mn@doi [Acta Physica Sinica]
  {{10.7498/aps.70.20210522}}, {70}, 163302

\bibitem[\protect\citeauthoryear{Juncher, Jorgensen  \& Helling}{Juncher
  et~al.}{2017}]{Juncher:2017}
Juncher D.,  Jorgensen U.~G.,   Helling C.,  2017, \mn@doi [A\&A]
  {10.1051/0004-6361/201629977}, 608, A70

\bibitem[\protect\citeauthoryear{Kirkpatrick}{Kirkpatrick}{2005}]{05Kixxxx.dwarfs}
Kirkpatrick J.~D.,  2005, \mn@doi [Annu. Rev. Astron. Astrophys.]
  {10.1146/annurev.astro.42.053102.134017}, 43, 195

\bibitem[\protect\citeauthoryear{Knowles \& Werner}{Knowles \&
  Werner}{1985}]{85KnWexx.ai}
Knowles P.~J.,  Werner H.-J.,  1985, \mn@doi [Chem. Phys. Lett.]
  {10.1016/0009-2614(85)80025-7}, 115, 259

\bibitem[\protect\citeauthoryear{Koput \& Peterson}{Koput \&
  Peterson}{2002}]{02KoPexx.CaOH}
Koput J.,  Peterson K.~A.,  2002, \mn@doi [J. Phys. Chem. A]
  {10.1021/jp026283u}, 106, 9595

\bibitem[\protect\citeauthoryear{Le~Roy \& Henderson}{Le~Roy \&
  Henderson}{2007}]{MLRpaper}
Le~Roy R.~J.,  Henderson R. D.~E.,  2007, \mn@doi [Mol. Phys.]
  {10.1080/00268970701241656}, 105, 663

\bibitem[\protect\citeauthoryear{Lee, Seto, Hirao, Bernath  \& Le~Roy}{Lee
  et~al.}{1999}]{EMO}
Lee E.~G.,  Seto J.~Y.,  Hirao T.,  Bernath P.~F.,   Le~Roy R.~J.,  1999,
  \mn@doi [J. Mol. Spectrosc.] {10.1006/jmsp.1998.7789}, 194, 197

\bibitem[\protect\citeauthoryear{Leininger \& Jeung}{Leininger \&
  Jeung}{1995}]{95LeJexx.CaH}
Leininger T.,  Jeung G.~H.,  1995, \mn@doi [J. Chem. Phys.] {10.1063/1.469581},
  103, 3942

\bibitem[\protect\citeauthoryear{Lemoine, Demuynck, Destombes  \&
  Davies}{Lemoine et~al.}{1988}]{88LeDeDe.MgH}
Lemoine B.,  Demuynck C.,  Destombes J.~L.,   Davies P.~B.,  1988, \mn@doi [J.
  Chem. Phys.] {10.1063/1.455188}, 89, 673

\bibitem[\protect\citeauthoryear{Leopold, Zink, Evenson, Jennings  \&
  Mizushima}{Leopold et~al.}{1986}]{86LeZiEv.MgH}
Leopold K.~R.,  Zink L.~R.,  Evenson K.~M.,  Jennings D.~A.,   Mizushima M.,
  1986, \mn@doi [J. Chem. Phys.] {10.1063/1.450445}, 84, 1935

\bibitem[\protect\citeauthoryear{Malik, Kitzmann, Mendonca, Grimm, Marleau,
  Linder, Tsai  \& Heng}{Malik et~al.}{2019}]{Malik:2019}
Malik M.,  Kitzmann D.,  Mendonca J.~M.,  Grimm S.~L.,  Marleau G.-D.,  Linder
  E.~F.,  Tsai S.-M.,   Heng K.,  2019, \mn@doi [AJ.]
  {10.3847/1538-3881/ab1084}, 157, 170

\bibitem[\protect\citeauthoryear{McWilliam \& Lambert}{McWilliam \&
  Lambert}{1988}]{88McLaxx.MgH}
McWilliam A.,  Lambert D.~L.,  1988, MNRAS, 230, 573

\bibitem[\protect\citeauthoryear{Meuwly \& Hutson}{Meuwly \&
  Hutson}{1999}]{99MeHuxx.methods}
Meuwly M.,  Hutson J.~M.,  1999, \mn@doi [J. Chem. Phys.] {{10.1063/1.478744}},
  {110}, 8338

\bibitem[\protect\citeauthoryear{Mostafanejad \& Shayesteh}{Mostafanejad \&
  Shayesteh}{2012}]{12MoShxx.MgH}
Mostafanejad M.,  Shayesteh A.,  2012, \mn@doi [Chemical Physics Letters]
  {10.1016/j.cplett.2012.08.056}, 551, 13

\bibitem[\protect\citeauthoryear{Olmstead}{Olmstead}{1908}]{08Olmste.CaH}
Olmstead C.~M.,  1908, \mn@doi [ApJ] {10.1086/141525}, 27, 66

\bibitem[\protect\citeauthoryear{Prascher, Woon, Peterson, Dunning  \&
  Wilson}{Prascher et~al.}{2011}]{11PaWoPe.ai}
Prascher B.~P.,  Woon D.~E.,  Peterson K.~A.,  Dunning T.~H.,   Wilson A.~K.,
  2011, \mn@doi [Theor. Chem. Acc.] {10.1007/s00214-010-0764-0}, 128, 69

\bibitem[\protect\citeauthoryear{Sakamoto, White, Kawaguchi, Ohishi, Usuda  \&
  Hasegawa}{Sakamoto et~al.}{1998}]{98SaWhKa.CaH}
Sakamoto S.,  White G.~J.,  Kawaguchi K.,  Ohishi M.,  Usuda K.~S.,   Hasegawa
  T.,  1998, \mn@doi [MNRAS] {10.1046/j.1365-8711.1998.02080.x}, 301, 872

\bibitem[\protect\citeauthoryear{Sedaghati et~al.,}{Sedaghati
  et~al.}{2017}]{17SeBoMa.TiO}
Sedaghati E.,  et~al., {2017}, \mn@doi [Nature] {10.1038/nature23651}, {549},
  238+

\bibitem[\protect\citeauthoryear{Semenov, Yurchenko  \& Tennyson}{Semenov
  et~al.}{2017}]{jt656}
Semenov M.,  Yurchenko S.~N.,   Tennyson J.,  2017, \mn@doi [J. Mol.
  Spectrosc.] {10.1016/j.jms.2016.11.004}, 330, 57

\bibitem[\protect\citeauthoryear{Shayesteh \& Bernath}{Shayesteh \&
  Bernath}{2011}]{11ShBexx.MgH}
Shayesteh A.,  Bernath P.~F.,  2011, \mn@doi [J. Chem. Phys.]
  {10.1063/1.3631341}, 135, 094308

\bibitem[\protect\citeauthoryear{Shayesteh, Appadoo, Gordon, Le~Roy  \&
  Bernath}{Shayesteh et~al.}{2004a}]{04ShApGo.MgH}
Shayesteh A.,  Appadoo D. R.~T.,  Gordon I.,  Le~Roy R.~J.,   Bernath P.~F.,
  2004a, \mn@doi [J. Chem. Phys.] {10.1063/1.1724821}, 120, 10002

\bibitem[\protect\citeauthoryear{Shayesteh, Walker, Gordon, Appadoo  \&
  Bernath}{Shayesteh et~al.}{2004b}]{04ShWaGo.CaH}
Shayesteh A.,  Walker K.~A.,  Gordon I.,  Appadoo D. R.~T.,   Bernath P.~F.,
  2004b, \mn@doi [J. Mol. Struct.] {10.1016/j.molstruc.2003.11.001}, 695, 23

\bibitem[\protect\citeauthoryear{Shayesteh, Henderson, Le~Roy  \&
  Bernath}{Shayesteh et~al.}{2007}]{07ShHeLe.MgH}
Shayesteh A.,  Henderson R. D.~E.,  Le~Roy R.~J.,   Bernath P.~F.,  2007,
  \mn@doi [J. Phys. Chem. A] {10.1021/jp075704a}, 111, 12495

\bibitem[\protect\citeauthoryear{Shayesteh, Ram  \& Bernath}{Shayesteh
  et~al.}{2013}]{13ShRaBe.CaH}
Shayesteh A.,  Ram R.~S.,   Bernath P.~F.,  2013, \mn@doi [J. Mol. Spectrosc.]
  {10.1016/j.jms.2013.04.009}, 288, 46

\bibitem[\protect\citeauthoryear{Shayesteh, Alavi, Rahman  \&
  Gharib-Nezhad}{Shayesteh et~al.}{2017}]{17ShAlRa.CaH}
Shayesteh A.,  Alavi S.~F.,  Rahman M.,   Gharib-Nezhad E.,  2017, \mn@doi
  [Chem. Phys. Lett.] {10.1016/j.cplett.2016.11.020}, 667, 345

\bibitem[\protect\citeauthoryear{Skokov, Peterson  \& Bowman}{Skokov
  et~al.}{1999}]{99SkPeBo.methods}
Skokov S.,  Peterson K.~A.,   Bowman J.~M.,  1999, \mn@doi [Chem. Phys. Lett.]
  {{10.1016/S0009-2614(99)00996-3}}, {312}, 494

\bibitem[\protect\citeauthoryear{Skory, Weck, Stancil  \& Kirby}{Skory
  et~al.}{2003}]{03SkWeSt.MgH}
Skory S.,  Weck P.~F.,  Stancil P.~C.,   Kirby K.,  2003, \mn@doi [ApJS]
  {10.1086/376834}, 148, 599

\bibitem[\protect\citeauthoryear{Snellen}{Snellen}{2014}]{14Snellen}
Snellen I.,  2014, \mn@doi [Phil. Trans. Royal Soc. London A]
  {10.1098/rsta.2013.0075}, 372, 20130075

\bibitem[\protect\citeauthoryear{Szidarovszky \& Cs\'{a}sz\'{a}r}{Szidarovszky
  \& Cs\'{a}sz\'{a}r}{2015}]{15SzCsxx.MgH}
Szidarovszky T.,  Cs\'{a}sz\'{a}r A.~G.,  2015, \mn@doi [J. Chem. Phys.]
  {10.1063/1.4904858}, 142, 014103

\bibitem[\protect\citeauthoryear{Tennyson \& Yurchenko}{Tennyson \&
  Yurchenko}{2017}]{jt626}
Tennyson J.,  Yurchenko S.~N.,  2017, \mn@doi [Int. J. Quantum Chem.]
  {10.1002/qua.25190}, 117, 92

\bibitem[\protect\citeauthoryear{Tennyson, Lodi, McKemmish  \&
  Yurchenko}{Tennyson et~al.}{2016}]{jt632}
Tennyson J.,  Lodi L.,  McKemmish L.~K.,   Yurchenko S.~N.,  2016, J. Phys. B:
  At. Mol. Opt. Phys., 49, 102001

\bibitem[\protect\citeauthoryear{Tennyson et~al.,}{Tennyson
  et~al.}{2020}]{jt810}
Tennyson J.,  et~al., 2020, \mn@doi [J. Quant. Spectrosc. Radiat. Transf.]
  {10.1016/j.jqsrt.2020.107228}, 255, 107228

\bibitem[\protect\citeauthoryear{T\'obi\'as, Furtenbacher, Tennyson  \&
  Cs\'asz\'ar}{T\'obi\'as et~al.}{2019}]{jt750}
T\'obi\'as R.,  Furtenbacher T.,  Tennyson J.,   Cs\'asz\'ar A.~G.,  2019,
  \mn@doi [Phys. Chem. Chem. Phys.] {10.1039/c8cp05169k}, 21, 3473

\bibitem[\protect\citeauthoryear{Tomkin \& Lambert}{Tomkin \&
  Lambert}{1980}]{80ToLaxx.MgH}
Tomkin J.,  Lambert D.~L.,  1980, \mn@doi [ApJ] {10.1086/157697}, 235, 925

\bibitem[\protect\citeauthoryear{Wallace, Hinkle, Li  \& Bernath}{Wallace
  et~al.}{1999}]{99WaHiLi.MgH}
Wallace L.,  Hinkle K.,  Li G.,   Bernath P.,  1999, \mn@doi [ApJ]
  {10.1086/307798}, 524, 454

\bibitem[\protect\citeauthoryear{Wang, Tennyson  \& Yurchenko}{Wang
  et~al.}{2020}]{jt790}
Wang Y.,  Tennyson J.,   Yurchenko S.~N.,  2020, \mn@doi [Atoms]
  {10.3390/atoms8010007}, 8, 7

\bibitem[\protect\citeauthoryear{Watanabe, Yoneyama, Uchida, Kobayashi,
  Matsushima, Moriwaki  \& Ross}{Watanabe et~al.}{2016}]{16WaYoUc.CaH}
Watanabe K.,  Yoneyama N.,  Uchida K.,  Kobayashi K.,  Matsushima F.,  Moriwaki
  Y.,   Ross S.~C.,  2016, \mn@doi [Chem. Phys. Lett.]
  {10.1016/j.cplett.2016.05.033}, 657, 1

\bibitem[\protect\citeauthoryear{Watanabe, Tani, Kobayashi, Moriwaki  \&
  Ross}{Watanabe et~al.}{2018}]{18WaTaKo.CaH}
Watanabe K.,  Tani I.,  Kobayashi K.,  Moriwaki Y.,   Ross S.~C.,  2018,
  \mn@doi [Chem. Phys. Lett.] {10.1016/j.cplett.2018.08.055}, 710, 11

\bibitem[\protect\citeauthoryear{Weck, Stancil  \& Kirby}{Weck
  et~al.}{2003a}]{03WeStKi.CaH}
Weck P.~F.,  Stancil P.~C.,   Kirby K.,  2003a, \mn@doi [J. Chem. Phys.]
  {10.1063/1.1573181}, 118, 9997

\bibitem[\protect\citeauthoryear{Weck, Schweitzer, Stancil, Hauschildt  \&
  Kirby}{Weck et~al.}{2003b}]{03WeScSt.MgHline}
Weck P.~F.,  Schweitzer A.,  Stancil P.~C.,  Hauschildt P.~H.,   Kirby K.,
  2003b, \mn@doi [ApJ] {10.1086/344722}, 582, 1059

\bibitem[\protect\citeauthoryear{Werner \& Knowles}{Werner \&
  Knowles}{1985}]{85WeKnxx.ai}
Werner H.-J.,  Knowles P.~J.,  1985, \mn@doi [J. Chem. Phys.]
  {10.1063/1.448627}, 82, 5053

\bibitem[\protect\citeauthoryear{Werner, Knowles, Knizia, Manby  \&
  Sch\"utz}{Werner et~al.}{2012}]{MOLPRO}
Werner H.-J.,  Knowles P.~J.,  Knizia G.,  Manby F.~R.,   Sch\"utz M.,  2012,
  \mn@doi [WIREs Comput. Mol. Sci.] {10.1002/wcms.82}, 2, 242

\bibitem[\protect\citeauthoryear{Werner et~al.,}{Werner
  et~al.}{2020}]{Molpro:JCP:2020}
Werner H.-J.,  et~al., 2020, \mn@doi [J. Chem. Phys.] {10.1063/5.0005081}, 152,
  144107

\bibitem[\protect\citeauthoryear{Yadin, Vaness, Conti, Hill, Yurchenko  \&
  Tennyson}{Yadin et~al.}{2012}]{jt529}
Yadin B.,  Vaness T.,  Conti P.,  Hill C.,  Yurchenko S.~N.,   Tennyson J.,
  2012, MNRAS, 425, 34

\bibitem[\protect\citeauthoryear{Yong, Lambert  \& Ivans}{Yong
  et~al.}{2003}]{03DaYDaL.MgH}
Yong D.,  Lambert D.~L.,   Ivans I.~I.,  2003, ApJ, 599, 1357

\bibitem[\protect\citeauthoryear{Yurchenko, Tennyson, Bailey, Hollis  \&
  Tinetti}{Yurchenko et~al.}{2014}]{jt572}
Yurchenko S.~N.,  Tennyson J.,  Bailey J.,  Hollis M. D.~J.,   Tinetti G.,
  2014, \mn@doi [Proc. Nat. Acad. Sci.] {10.1073/pnas.1324219111}, 111, 9379

\bibitem[\protect\citeauthoryear{Yurchenko, Lodi, Tennyson  \&
  Stolyarov}{Yurchenko et~al.}{2016}]{jt609}
Yurchenko S.~N.,  Lodi L.,  Tennyson J.,   Stolyarov A.~V.,  2016, \mn@doi
  [Comput. Phys. Commun.] {10.1016/j.cpc.2015.12.021}, 202, 262

\bibitem[\protect\citeauthoryear{Yurchenko, Al-Refaie  \& Tennyson}{Yurchenko
  et~al.}{2018}]{jt708}
Yurchenko S.~N.,  Al-Refaie A.~F.,   Tennyson J.,  2018, \mn@doi [A\&A]
  {10.1051/0004-6361/201732531}, 614, A131

\bibitem[\protect\citeauthoryear{Zhang \& Steimle}{Zhang \&
  Steimle}{2014}]{14ZhStxx.MgH}
Zhang R.,  Steimle T.~C.,  {2014}, \mn@doi [ApJ]
  {{10.1088/0004-637X/781/1/51}}, {781}, 51

\bibitem[\protect\citeauthoryear{Zink, Jennings, Evenson  \& Leopold}{Zink
  et~al.}{1990}]{90ZiJeEv.MgH}
Zink L.~R.,  Jennings D.~A.,  Evenson K.~M.,   Leopold K.~R.,  1990, \mn@doi
  [ApJ] {10.1086/185796}, 359, L65

\bibitem[\protect\citeauthoryear{Ziurys, Barclay  \& Anderson}{Ziurys
  et~al.}{1993}]{93ZiBaAn.MgH}
Ziurys L.~M.,  Barclay W.~L.,   Anderson M.~A.,  1993, \mn@doi [ApJ]
  {10.1086/186690}, 402, L21

\makeatother
\end{thebibliography}

\bsp	
\label{lastpage}
\end{document}